\documentclass[journal,twoside,web]{ieeecolor}
\usepackage{tmi}
\usepackage{cite}
\usepackage{amsmath,amssymb,amsfonts}
\usepackage{algorithmic}
\usepackage{graphicx}
\usepackage{textcomp}

\usepackage{latexsym}
\usepackage{hyperref}
\usepackage{booktabs}
\usepackage{pgfplots}
\usepackage{multirow}
\usepackage{float}
\usepackage{mathrsfs}
\usepackage[ruled]{algorithm2e}
\usepackage{etoolbox}

\usepackage{soul}
\usepackage{xcolor}
\usepackage{tabularray}
\usepackage{lipsum}
\usepackage{xr}
\usepackage[switch]{lineno}
\usepackage{comment}
\usepackage{subfigure}
\usepackage{hhline}

\definecolor{TMIblue}{HTML}{0086D9}
\makeatletter
\renewcommand{\@cite}[1]{\textcolor{TMIblue}{[#1]}}
\makeatother

\providecommand{\Leireftb}[1]{Table~\ref{#1}}
\providecommand{\Leireffig}[1]{Fig.~\ref{#1}}
\providecommand{\Leireffigure}[1]{Figure~\ref{#1}}

\providecommand{\cite}[1]{\cite{#1}}

\def\BibTeX{{\rm B\kern-.05em{\sc i\kern-.025em b}\kern-.08em
    T\kern-.1667em\lower.7ex\hbox{E}\kern-.125emX}}
\begin{document}

\title{Towards Enabling Cardiac Digital Twins of Myocardial Infarction Using Deep Computational Models for Inverse Inference}
\author{Lei Li, Julia Camps, Zhinuo (Jenny) Wang, Marcel Beetz, Abhirup Banerjee, Blanca Rodriguez, and Vicente Grau
\thanks{Corresponding author: Lei Li (e-mail: lei.li@eng.ox.ac.uk). 
This work was supported by the CompBioMed 2 Centre of Excellence in Computational Biomedicine (European Commission Horizon 2020 research and innovation programme, grant agreement No. 823712).
L. Li was partially supported by the SJTU 2021 Outstanding Doctoral Graduate Development Scholarship.
A. Banerjee was supported by the Royal Society Grant No. URF{\textbackslash}R1{\textbackslash}221314.
A. Banerjee and V. Grau were partially supported by the British Heart Foundation Project under Grant PG/20/21/35082.}
\thanks{Lei Li, Marcel Beetz, Abhirup Banerjee, and Vicente Grau are with the Department of Engineering Science, University of Oxford, Oxford, UK.}
\thanks{Julia Camps, Zhinuo (Jenny) Wang, and Blanca Rodriguez are with the Department of Computer Science, University of Oxford, Oxford, UK.}
}

\maketitle

\begin{abstract}
Cardiac digital twins (CDTs) have the potential to offer individualized evaluation of cardiac function in a non-invasive manner, making them a promising approach for personalized diagnosis and treatment planning of myocardial infarction (MI).
The inference of accurate myocardial tissue properties is crucial in creating a reliable CDT of MI.
In this work, we investigate the feasibility of inferring myocardial tissue properties from the electrocardiogram (ECG) within a CDT platform. 
The platform integrates multi-modal data, such as cardiac MRI and ECG, to enhance the accuracy and reliability of the inferred tissue properties.
We perform a sensitivity analysis based on computer simulations, systematically exploring the effects of infarct location, size, degree of transmurality, and electrical activity alteration on the simulated QRS complex of ECG, to establish the limits of the approach. 
We subsequently present a novel deep computational model, comprising a dual-branch variational autoencoder and an inference model, to infer infarct location and distribution from the simulated QRS. 
The proposed model achieves mean Dice scores of $ 0.457 \pm 0.317 $ and $ 0.302 \pm 0.273 $ for the inference of left ventricle scars and border zone, respectively.
The sensitivity analysis enhances our understanding of the complex relationship between infarct characteristics and electrophysiological features.
The \emph{in silico} experimental results show that the model can effectively capture the relationship for the inverse inference, with promising potential for clinical application in the future.
The code will be released publicly once the manuscript is accepted for publication.
\end{abstract}


\section{Introduction}
	
Myocardial infarction (MI) is a major cause of mortality and disability worldwide \cite{journal/lancet/john2012}.
Assessment of myocardial viability is essential in the diagnosis and treatment management for patients suffering from MI.
In particular, the location and distribution of myocardial scars provide important information for patient diagnosis and treatment.
Late gadolinium enhancement (LGE) magnetic resonance imaging (MRI) has been widely used to characterize myocardial scars \cite{journal/MedIA/li2023}.
However, the incorporation of LGE into MRI examination prolongs scan time and has potential side effects \cite{journal/Radiology/ordovas2011}. 
Recent studies have tried to delineate scars using non-enhanced MRI, with promising preliminary results \cite{journal/Radiology/zhang2019,journal/MedIA/xu2020a}. 
Alternatively, the electrocardiogram (ECG) can be used to reveal abnormalities related in electrophysiology post-MI \cite{journal/NEJM/zimetbaum2003}.
For example, ST-segment elevation and T-wave inversion are commonly used indicators of cardiac remodeling associated with different stages of MI \cite{journal/JE/hanna2011}.
In contrast, QRS patterns have received less attention in the literature, though they also provide valuable information about the extent and location of myocardial damage in the post-MI \cite{conf/FIMH/li2023}.
It is still partially unclear how QRS abnormalities reflect MI characteristics, such as location, size, transmural extent, and cardiac electrical activity alterations.
Therefore, it is highly desirable to investigate their relationships and thus better understand the diagnostic and prognostic value of QRS abnormalities for MI.

\begin{figure*}[t]\center
 \includegraphics[width=0.82\textwidth]{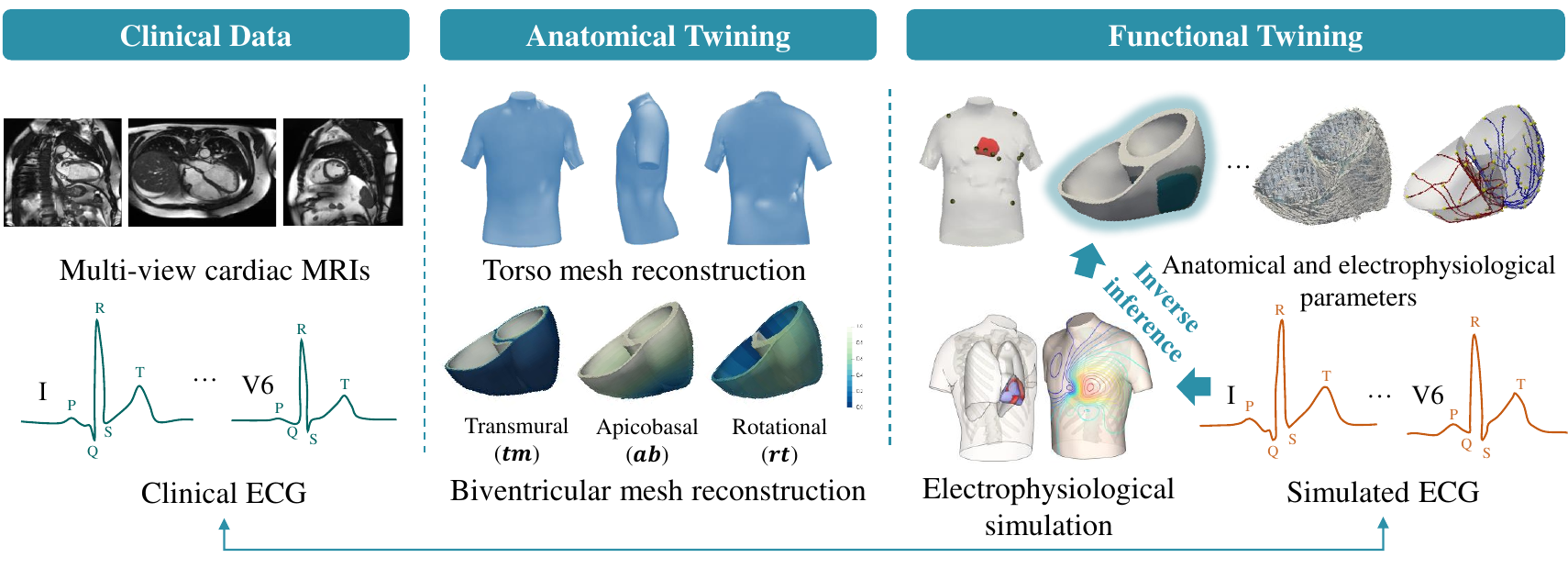}\\[-2ex]
   \caption{The cardiac digital twin (CDT) generation workflow combining cine cardiac magnetic resonance images (MRIs) and electrocardiogram (ECG). 
   Here, the anatomical twinning personalizes the geometrical model, while functional twinning personalizes the electrophysiological model.
   The anatomical and electrophysiological parameters include electrode positions, myocardial infarction (MI) distribution, ventricular muscle fiber orientation, Purkinje system, etc. 
   Our goal is to solve the inverse problem for inferring the infarct location map (highlighted via the glow effect) from simulated QRS.} 
\label{fig:intro:CDT}
\end{figure*}

Cardiac ``digital twin" (CDT) technology can create virtual models of the heart combining cardiac images, ECG, and other subject-specific information \cite{journal/EHJ/corral2020}. 
It allows clinicians to visualize and analyze the structure, function, and electrical activity of the heart in real-time, providing valuable insights into the underlying mechanisms of MI \cite{journal/NC/arevalo2016}.
As \Leireffig{fig:intro:CDT} shows, CDT workflows usually involve two stages, namely anatomical and functional twinnings, which present various challenges \cite{journal/MedIA/gillette2021}.
The anatomical twinning stage involves the segmentation of cardiac images, reconstruction of the 3D geometry of the heart, and the identification and extraction of relevant anatomical structures. 
It is complicated by the variability in the cardiac anatomy across individuals, as well as by imaging artifacts and noise.
At the functional twinning stage, the main challenge is to solve the inverse problem of electrocardiography, i.e. estimating cardiac activity from the measured ECG, which is inherently ill-posed, meaning that multiple solutions can lead to the same observed data.
This is complicated by the limitations of ECG recordings, which are sparse, noisy, and subject to substantial uncertainties.

In this work, we develop a deep computational model for the inverse inference of post-MI with different properties, varying the infarct location, size, and transmural extent. 
We first conduct a sensitivity analysis to investigate the relationship between QRS abnormalities and infarct characteristics in post-MI. 
This analysis provides insights into how variations in QRS signals are associated with specific infarct properties, informing the subsequent inference process.
We then propose an end-to-end inverse inference framework that leverages a multi-modal variational autoencoder (VAE) in conjunction with an inference model.
The framework can efficiently combine the anatomical properties from cine MRI and electrophysiological information from simulated QRS.
This study provides an integrated and personalized perspective that incorporates the features from multi-modal data to predict tissue properties of post-MI, enabling the construction of a CDT platform.
To the best of our knowledge, this is the first deep learning based computational model that addresses the inverse inference of MI with diverse characteristics while incorporating a comprehensive sensitivity analysis.
The main contributions of this work include:
\begin{itemize}
  \item {We develop a novel deep computational model for the inverse inference of infarct regions from simulated QRS and point cloud reconstructed from non-enhanced MRI.}
  \item {We perform a comprehensive sensitivity analysis to investigate the relationship between QRS abnormalities and infarct characteristics in post-MI.}
  \item {We utilize a unified coordinate system to consistently represent the ventricles and infarct characteristics, thus mitigating the impact of inter-subject anatomical variations.}  
  \item {We prove the feasibility of inferring myocardial tissue properties from multi-modal data to create a reliable CDT platform for personalize medicine.}
\end{itemize}

\section{Related Work}

\subsection{Myocardial Infarction Detection from MRI or ECG}

LGE MRI has been widely used to visualize LV scarring area, while T2-weighted MRI is employed to depict border zone \cite{journal/MedIA/qiu2023}.
However, automatic segmentation of LV scarring area/ border zone from MRI could be quite challenging, due to low image quality and the inherent variability in the appearance of pathological tissue.
A few studies have attempted to address these challenges by combining multi-sequence MRI for scar and border zone segmentation \cite{journal/MedIA/wang2022,journal/TMI/ding2023}.
The majority have primarily focused on LV scar segmentation using conventional methods, such as thresholding \cite{journal/MedIA/karim2016}, region growing \cite{journal/MRM/tao2010}, fuzzy clustering \cite{conf/CC/baron2008}, continuous max-flow \cite{journal/TMI/ukwatta2015}, and graph-cuts \cite{journal/TMI/rajchl2013}. 
Recent advancements in deep learning have yielded promising results, offering more efficient and accurate models for LV scar segmentation from LGE MRI \cite{journal/MRMPBM/moccia2019,journal/CDHJ/popescu2022}. 
Most work adopted a two-stage model by extracting myocardium as a prior and then performing a pathology segmentation \cite{journal/MedIA/li2023}.
Nevertheless, LGE MRI could be cost-prohibitive and potentially pose risks to certain patients.
Therefore, several studies have explored employing contrast agent-free cine MRI for infarct segmentation via motion traction or LGE MRI synthesis \cite{journal/Radiology/zhang2019,journal/MedIA/xu2020a}.
Instead of relying on imaging data for precise infarct area localization, an alternative approach is to use ECG to perform the preliminary coarse location classification of the infarct area \cite{journal/JC/wieslander2013,journal/PRL/baloglu2019,journal/CMPB/xiong2021}.
Ghimire \textit{et al.} \cite{journal/TMI/ghimire2019} employed variational approximation to reconstruct transmural action potential from CT and 120-lead ECG and then extracted scarring area via thresholding the activation time.
To the best of our knowledge, there has been no prior research that combines non-enhanced MRI data and ECG for the localization of MI with different characteristics.

\subsection{Integration of Cardiac Images with Non-Imaging Information}

The fusion of information from multiple modalities, such as anatomical images, ECG signals, and clinical metadata, holds the potential to provide a more comprehensive understanding of the underlying cardiac electrical activity \cite{journal/MedIA/li2023survey}.
For instance, one can align CT and electroanatomical mapping (EAM) data via landmark-based registration for combining anatomical and electrical information to guide ablation procedures \cite{journal/PRO/brett2021}.
The fusion can be achieved simply based on a spatial registration, as EAM data inherently encodes spatial information about the electrical activity within the heart. 
However, the data representation of non-imaging data is normally different from imaging data, and thus introduces additional challenges for multi-modal data integration.
To solve this, several studies have leveraged contrastive learning techniques to establish correspondences between imaging and non-imaging pairs within randomly sampled batches \cite{conf/ICML/radford2021,conf/ICML/jia2021}.
It is important to note that these approaches assume a prerequisite alignment of input modalities, which may be invalid in some scenarios.
Alternatively, one can construct the anatomical mesh from imaging data, followed by the mapping of non-imaging information to individual vertices of the mesh \cite{conf/STACOM/meister2020}.
Recent advancements in multi-modal fusion methods fundamentally aim to integrate multi-modal data into a global feature space, allowing for a uniform representation of the integrated information \cite{conf/WWWC/khattar2019,conf/MICCAI/aguila2023}. 
For instance, Li \textit{et al.} \cite{conf/STACOM/li2023} employed multi-model representation learning to combine anatomical images and ECG signals for the inference of ventricular activation properties.
The integration of anatomical images and ECG signals in a unified feature space enables the inverse inference of critical parameters related to ventricular activation.  

\subsection{Solving the Electrocardiography Inverse Problem}

Estimation of electrical activity inside the heart from body surface potentials or ECG is known as an inverse problem of electrocardiography.
To solve the inverse problem, state-of-the-art approaches can be coarsely separated into two kinds: deterministic and probabilistic methods \cite{book/SSBM/kaipio2006}.
Deterministic approaches in cardiac electrophysiology involve minimizing a cost function that quantifies the discrepancy between the observed data and the model predictions.
For robust inverse, spatial and/ or temporal regularizations have been widely used \cite{journal/TMI/yu2015,journal/FP/karoui2018}. 
However, deterministic optimization provides single-point estimates of model parameters without considering measurement data uncertainty. 
Probabilistic methods rely on Bayesian inference theory and numerical techniques to generate posterior distributions for the model parameters \cite{conf/MICCAI/xu2015,journal/MedIA/camps2021}.
They can incorporate prior knowledge into the parameter estimation with uncertainty, which can be used to guide decision-making and assess the robustness of the results \cite{journal/TMI/ghimire2019}.
Nevertheless, conventional probabilistic methods are usually computationally expensive, as repeated numerical simulations are required to generate samples for the posterior distribution.
Recently, deep learning based probabilistic methods have emerged as an alternative to conventional methods for modeling complex dynamics of cardiac electrical activity. 
They can leverage deep neural networks to approximate the posterior distribution of the model parameters or latent variables, providing faster and more accurate approximations.
For example, Meister \textit{et al.} \cite{conf/STACOM/meister2020} employed graph convolutional neural networks to estimate the depolarization patterns in the myocardium with scars.
Bacoyannis \textit{et al.} \cite{journal/EP/bacoyannis2021} reconstructed activation patterns of the myocardium with various local wall thicknesses, as thin walls indicate infarct regions.
However, with regards to different post-MI scenarios, the inverse inference of electrophysiological heterogeneity in the infarct regions has not been fully investigated yet.

Given the ill-posed nature of the inverse problem, it is imperative conduct thorough and rigorous validation to ensure its clinical acceptance.
The majority of the validation studies employed computer models for \emph{in silico} evaluation \cite{journal/TBME/schuler2021,journal/FiP/kalinin2019,journal/TMI/ghimire2019} and/ or performed evaluation on the basis of torso-tank experiments with isolated animal hearts, such as rabbit \cite{journal/AJPHCP/zhang2005}, canine \cite{journal/MBEC/cluitmans2018,journal/TMI/jiang2022}, and swine \cite{conf/STACOM/meister2021,journal/Sensors/chen2022}.
There are only few studies evaluated using in-vivo human subjects, which however usually employed non-simultaneous invasive recordings, such as electrograms \cite{journal/EP/pezzuto2021} or bipolar voltages maps \cite{conf/MICCAI/xu2014,journal/TMI/ghimire2019}.

\section{Methodology}

\begin{figure*}[t]\center
 \includegraphics[width=1\textwidth]{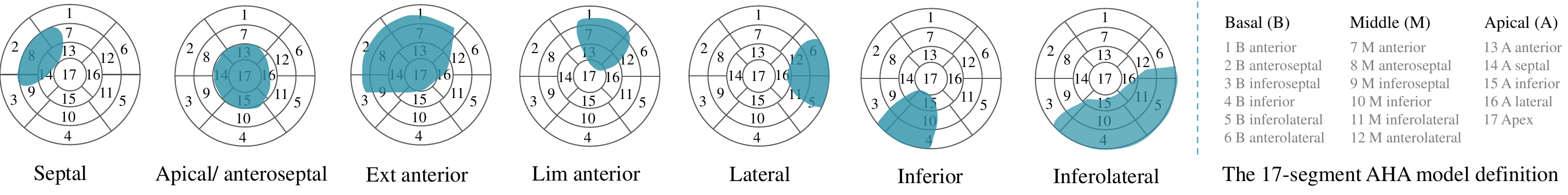}\\[-2ex]
   \caption{The seven infarct locations defined on the 17-segment American Heart Association (AHA) model. 
   The selection of the seven locations is referring to \cite{journal/CCR/nikus2014}. Ext: extensive; Lim: limited.}
\label{fig:method:17AHA_MI_location}
\end{figure*}

\begin{figure*}[t]\center
 \includegraphics[width=0.78\textwidth]{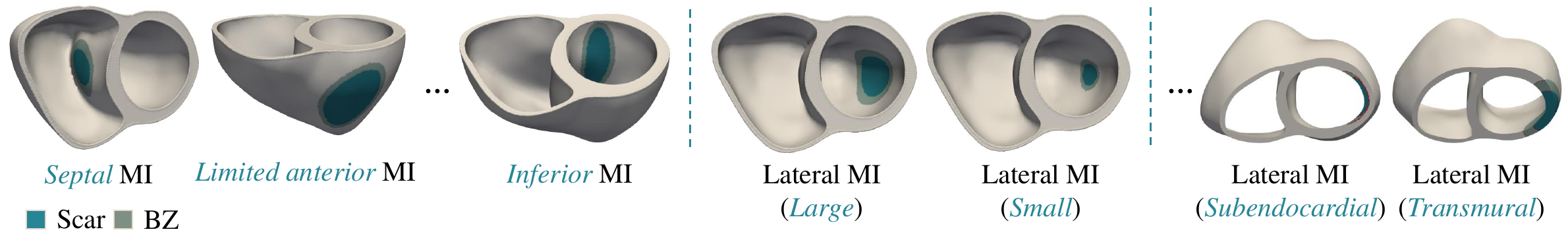}\\[-2ex]
   \caption{Illustration of several post-MI scenarios, including different infarct locations, sizes, and transmural extents. Here, scars refer to the area of damaged or dead heart muscle tissue that has been replaced by non-functional fibrous tissue, while the border zone (BZ) is the area surrounding the scar tissue where there may be some remaining damaged heart muscle tissue that is not yet fully scarred.}
\label{fig:method:MI_examples}
\end{figure*}

\subsection{Anatomical Twinning: Mesh Reconstruction} \label{method:mesh generation}

At the anatomical twinning stage, we reconstruct a subject-specific 3D torso-biventricular tetrahedral mesh from multi-view cardiac MRIs \cite{journal/PTRSA/banerjee2021,conf/EMBC/smith2022}.
Specifically, for the biventricular reconstruction, we first use a deep learning based ventricle segmentation from long- and short-axis cardiac cine MRIs at the end-diastolic (ED) phase and thus obtain sparse 3D contours.
We then perform a misalignment correction based on the intensity and contour information coupled with a statistical shape model, followed by a surface mesh reconstruction and volumetric tetrahedral mesh generation.
We utilize a two-step automated framework for the torso reconstruction, and the locations of the ECG electrodes (I, II, V1-V6, LA, RA, LL, RL) are measured from the personalized 3D torso mesh.
To ensure a symmetric, consistent, and intuitive biventricular representation across various geometries, we project the biventricular mesh into a consistent biventricular coordinates (Cobiveco) system \cite{journal/MedIA/schuler2021}.
The Cobiveco system is defined by $(tm, ab, rt, tv)$, which correspond to transmural, apicobasal, rotational, and transventricular coordinates, respectively.
The reader is referred to the anatomical twinning stage of \Leireffig{fig:intro:CDT} for the illustration of Cobiveco ($tv$ is excluded there).
We represent infarct areas in the myocardium as an ellipse with radii $r_{tm}$, $r_{ab}$, and $r_{rt}$ as follows,
\begin{equation}
	\frac {({tm}_i - {tm}_0)^2} {{r_{tm}}^2} + \frac {({ab}_i - {ab}_0)^2} {{r_{ab}}^2} + \frac {({rt}_i - {rt}_0)^2} {{r_{rt}}^2} \leq 1,
\end{equation}
where $({tm}_0, {ab}_0, {rt}_0)$ is the center coordinate of the scar region.


We consider different post-MI scenarios, including seven locations, two transmural extents, two different sizes, and two different cardiac electrical activity alterations.
As \Leireffig{fig:method:17AHA_MI_location} shows, one can define the infarct areas consistently in the 17-segment American Heart Association (AHA) map \cite{journal/CCR/nikus2014}, enabling the study of the effects of MI properties at a population level.
Note that in this study, we only consider the scars in the left ventricle (LV), as the majority of clinically significant myocardial scars present there \cite{journal/EHJ/thygesen2019}. 
The LV region is defined in Cobiveco as $tv = 0 \vee (tv = 1 \wedge rt > \frac{2}{3})$ to include the whole septum.
For the comparison of different infarct sizes and cardiac electrical activity alterations, we only report on lateral MI as an illustrative case.
We simulate infarct at seven different locations and one smaller size on lateral MI, each with two levels of transmural extent, and one scenario with a slower CV on transmural large lateral MI, resulting in a total of 17 post-MI scenarios for each subject.
\Leireffigure{fig:method:MI_examples} provides several examples of our experimental scenarios.

\subsection{Functional Twinning: Forward Electrophysiological Simulation} \label{method:simulation}

At the functional twinning stage, we simulate cardiac electrophysiology via an efficient orthotropic Eikonal model \cite{journal/MedIA/camps2021}, which incorporates a human-based Purkinje system into the formulation of the activation times of root nodes (RNs), i.e., the sites of earliest activation.
The simulation is performed on the Cobiveco mesh, solving,
\begin{equation}\label{eq:eikonal_eq}
\left\{
\begin{aligned}
& \sqrt{\nabla^T t\mathcal{V}^2 \nabla t} = 1, \\
& t(\Gamma_0) = pk\left(\Gamma_0\right)-\min\left(pk\left(\Gamma_0\right)\right),
\end{aligned}
\right.
\end{equation}
where $\mathcal{V}$ are the orthogonal conduction velocities (CVs) of fibre, sheet (transmural), and sheet-normal directions, $t$ is the time at which the activation wavefront reaches each point in the mesh, $\Gamma_0$ is the set of RN locations, and $pk$ is a Purkinje-tree delay function from the His-bundle to every point. 
Therefore, the earliest activation time at the RNs is defined as their delay from the His-bundle through the Purkinje tree normalized by the earliest activation, such that the wavefront originates at $t = 0$ in one of the endocardial RNs. 
The QRS can be calculated from the activation time map (ATM) via a pseudo-ECG equation \cite{journal/CR/gima2002} for a 1D cable source with constant conductivity at a given electrode location $(x',y',z')$ as follows
\begin{equation}
\phi_e \left(x',y',z' \right) = \frac{a^2 \sigma_i}{4 \sigma_e} \int - \nabla V_m \cdot \left[ \nabla \frac{1}{r} \right] \,dx \,dy \,dz \ ,
\end{equation}
where $V_m$ is the transmembrane potential, $\nabla V_m$ is its spatial gradient, $r$ is the Euclidean distance from a given point $(x,y,z)$ to the electrode location, $a$ is a constant that depends on the fiber radius, and $\sigma_i$ and $\sigma_e$ are the intracellular and extracellular conductivities, respectively. 
The pseudo-ECG method can efficiently generate normalized ECG signals without significant loss of morphological information compared to the bidomain simulation \cite{journal/FiP/minchole2019}.

\begin{figure*}[t]\center
 \includegraphics[width=0.76\textwidth]{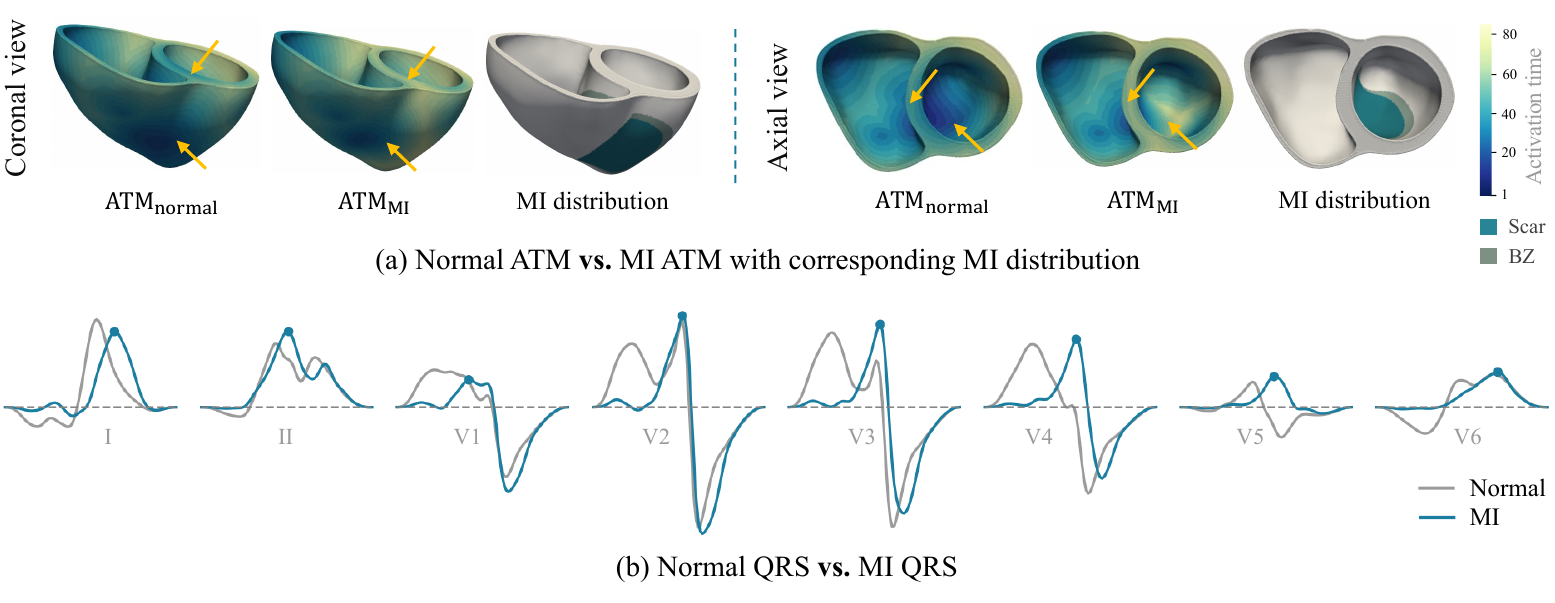}\\[-2ex]
   \caption{Illustration of regional alterations in ventricular activation when scars are present in the heart.
   Here, we employ the subject with transmural extensive anterior MI as an example, to compare its activation time map (ATM) and QRS with that of a corresponding healthy one.
   The arrows highlight the areas where ATM differs in MI and healthy cases.
   }
\label{fig:method:ATM MI vs. normal}
\end{figure*}

In modeling the effects of scars on the QRS, it is essential to consider the electrophysiological properties of the infarct regions, such as the slower CVs \cite{journal/Circulation/de1993}, which can lead to changes in the timing and amplitude of the ECG waveform and thus manifest as changes in the QRS.
Therefore, we vary the CVs of infarct and healthy myocardial areas during QRS simulation (see Sec.~\ref{exp:data info}).
As Fig.~\ref{fig:method:ATM MI vs. normal} shows, the ATM of MI patients presents slower electrical signal propagation compared to that of healthy ones, resulting in corresponding alteration in the simulated QRS morphology. 
The QRS signal was simulated for each MI scenario of each subject, establishing an ideal 1-to-1 relationship between the QRS signal and the MI scenario.
However, note that in cases where the assigned MI regions of two or more MI scenarios for a single subject have no distinguishable effects or produce identical effects on the final simulated QRS signals, this ideal 1-to-1 relationship may become invalid.

\subsection{Functional Twinning: Inverse Inference of Post-MI Properties} \label{method:computational model}

\begin{figure}[t]\center
 \includegraphics[width=0.48\textwidth]{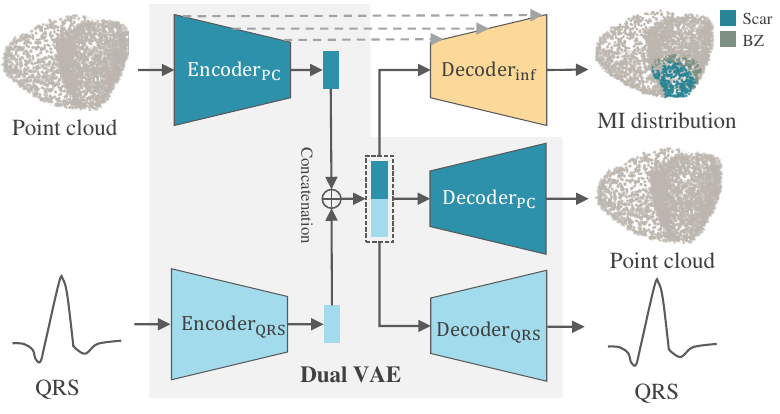}\\[-2ex]
   \caption{Deep computational model for the inverse inference of MI based on a dual variational autoencoder (VAE). 
 Note that the reconstructed point clouds (PCs) include both dense and sparse PCs and the simulated QRS includes 8 leads. For simplicity, the schematic of sparse PC is omitted, and only single lead is presented here.}
\label{fig:method:computation model}
\end{figure}

\begin{figure*}[t]\center
 \includegraphics[width=0.86\textwidth]{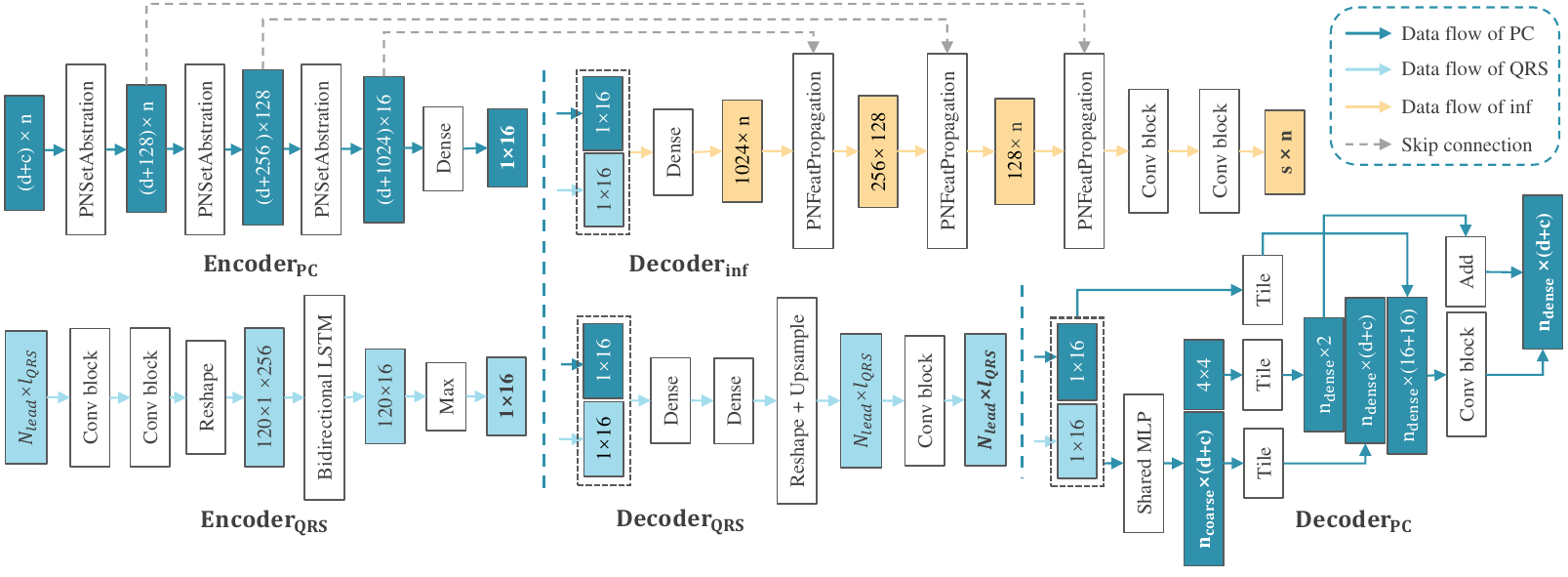}\\[-2ex]
   \caption{The network architecture of the proposed deep computational model.
   Here, $n$ is the number of nodes of input PC ($n=4096$), which includes three point coordinates ($d=3$) and four Cobiveco coordinates ($c=4$);
   $s$ is the number of categories of nodes which could be healthy, scar, and BZ ($s=3$);
   $n_\text{coarse}$ and $n_\text{dense}$ are the numbers of nodes in the coarse and dense output PCs, respectively ($n_\text{coarse}=1024$, $n_\text{dense}=4096$);
   $N_{lead}$ and $l_{QRS}$ refer the number of QRS leads and the unified length of QRS signals, respectively ($N_{lead}=8$, $l_{QRS}=512$).
   The latent space features of PC and QRS data have been concatenated and fed to the three decoders.}
\label{fig:method:network}
\end{figure*}

\Leireffigure{fig:method:computation model} provides an overview of the proposed deep computation model, consisting of a dual-branch VAE and an inference model.
The VAE captures both anatomical and electrophysiological features, while the inference model uses the latent space representation to predict scar and border zone location.
It offers a probabilistic framework that handles noise and uncertainty in ECG data, and the continuous latent space of VAE further provides an interpretable representation of MI properties, facilitating the inverse inference of MI \cite{conf/MICCAI/xu2014}.
\Leireffig{fig:method:network} depicts the network architecture.
One can see that the Encoder$_\text{PC}$ and Encoder$_\text{QRS}$ extract latent features from PC and QRS, separately.
The latent space features of PC and QRS have been concatenated, to gain a unified anatomical and electrophysiological representations.
To extract semantic features for the segmentation of scars/ BZ, the Encoder$_\text{PC}$ and Decoder$_\text{inf}$ are designed based on the architecture of PointNet++ \cite{journal/ANIP/qi2017}.
The design of Decoder$_\text{PC}$ is inspired by the work of Beetz \textit{et al.} \cite{journal/MedIA/beetz2023}.
We utilize bidirectional long short-term memory (LSTM) module \cite{journal/SR/zhu2019} in the Encoder$_\text{QRS}$ to capture temporal dependencies and sequential patterns within the QRS, which is inherently time-series data.

For the geometry reconstruction, we reconstruct coarse and dense point clouds (PCs) to simultaneously learn global shape and local anatomy of the ventricles \cite{journal/FiP/beetz2022,journal/MedIA/beetz2023}.
Therefore, the PC reconstruction loss function is defined as follows,
\begin{equation}
	\mathcal{L}^\text{rec}_\text{PC} = \sum_{i=1}^{K}\left(\mathcal{L}_{i,coarse}^\text{chamfer} + \alpha \mathcal{L}_{i,dense}^\text{chamfer}\right),
\end{equation}
where $K$ is the number of classes, $\alpha$ is the weight term between the two PCs, and $\mathcal{L}^\text{chamfer}$ is the chamfer distance between the input and reconstructed PCs.
To improve the fidelity and resemblance of the reconstructed $\text{Q}\hat{\text{R}}\text{S}$ to the original QRS, we minimize their mean-squared error (MSE) and dynamic time warping (DTW) distance \cite{journal/MedIA/camps2021},
\begin{equation}
	\mathcal{L}^\text{rec}_\text{QRS} = \mathcal{L}_\text{MSE}(\text{QRS}, \text{Q}\hat{\text{R}}\text{S}) + \mathcal{L}_\text{DTW}(\text{QRS}, \text{Q}\hat{\text{R}}\text{S}).
\end{equation} 
Finally, the loss function for training the dual-branch VAE is calculated as,
\begin{equation}
	\mathcal{L}_\text{VAE} = \lambda_\text{PC}\mathcal{L}^\text{rec}_\text{PC} + \lambda_\text{QRS}\mathcal{L}^\text{rec}_\text{QRS} + \lambda_\text{KL}\mathcal{L}^\text{KL},
\end{equation}
where $\lambda_\text{PC}$, $\lambda_\text{QRS}$, and $\lambda_\text{KL}$ are balancing parameters, and $\mathcal{L}^\text{KL}$ is the Kullback-Leibler (KL) divergence loss to mitigate the distance between the prior and posterior distributions of the latent space.
Here, the posterior distribution is presumed to follow a standard normal distribution, denoted as $\mathcal{N}(0, 1)$.

For the inference, we predict the infarct location based on the low-dimensional features learned from the VAE.
To alleviate the class-imbalance issue existed in the MI segmentation, we combine the cross-entropy (CE) loss and Dice loss,
\begin{equation}
	\mathcal{L}_\text{seg} = \mathcal{L}_\text{CE} + \lambda_\text{Dice}\mathcal{L}_\text{Dice},
\end{equation}
where $\lambda_\text{Dice}$ is a balancing parameter.
For realistic infarct shape, we further introduce a compactness loss,
\begin{equation}
	\mathcal{L}_\text{compact} = \frac{1}{N^{pre}} \sum_{i=1}^{N^{pre}} \frac{d_i^{pre} + d_i^{gd}} {d_{\max}^{gd}},
\end{equation}
where $N^{pre}$ is the total number of predicted MI points, $d_i^{pre}$ and $d_i^{gd}$ are the Euclidean distances from each predicted MI point $i$ to the center of predicted and ground truth MI, respectively,
and $d_{\max}^{gd}$ is the maximum Euclidean distance from ground truth MI points to their center.
We introduce two further constraints, to control infarct size and prevent scar from appearing in the right ventricle (RV) via,
\begin{equation}
	\mathcal{L}_\text{size} = \frac{N^{pre}-N^{gd}} {N^{gd}},
\end{equation}
\begin{equation}
	\mathcal{L}_\text{spa} = \frac{N^{pre}_{RV}} {N^{pre}},
\end{equation}
where $N^{gd}$ is the total number of ground truth infarct points, while $N^{pre}_{RV}$ is the number of predicted infarct points located in the RV, excluding the septum boundary.
Hence, the final inference loss is defined as,
\begin{equation}
	\mathcal{L}_\text{inf} = \lambda_\text{VAE}\mathcal{L}_\text{VAE} + \mathcal{L}_\text{seg} + \lambda_\text{compact}\mathcal{L}_\text{compact} + \lambda_\text{size}\mathcal{L}_\text{size} + \lambda_\text{spa}\mathcal{L}_\text{spa},
\end{equation}
where $\lambda_\text{compact}$, $\lambda_\text{size}$ and $\lambda_\text{spa}$ are balancing parameters.

\section{Experiments and Results}

\subsection{Materials}
\subsubsection{Dataset and Simulation Setup} \label{exp:data info}
We collected 49 subjects with paired 12-lead ECGs and multi-view cardiac MRIs from the UK Biobank study, under application number 40161 \cite{journal/EHJ/littlejohns2019}.
Each subject has a set of 17 simulated post-MI scenarios for sensitivity analysis. 
However, for the MI inference, only 16 scenarios were considered, as the scenario involving slower CV was excluded, resulting in a total of 784 experimental data (49 subjects $\times$ 16 scenarios).
The dataset was randomly divided into 34 training subjects, 5 validation subjects, and 10 test subjects.
The biventricular tetrahedral mesh for each subject was converted into PCs and then resampled into coarse and dense versions with 1,024 and 4,096 nodes, respectively.

Using simulated data allows for controlled exploration of the relationship between QRS abnormalities and MI characteristics and provides a known ground truth for inference evaluation, which is challenging with real data.
During the electrophysiology simulations, a fixed set of RN locations and CV values were utilized.
Specifically, the RNs were placed at seven specific homologous locations based on Cobiveco -- four in the LV and three in the RV. 
In the LV, they were situated in the mid-septum, basal-anterior paraseptal, and two mid-posterior locations, while in the RV, they were located in the mid-septum and two free wall regions \cite{journal/EP/cardone2016}.
The RN configuration was consistent in healthy and MI subjects across different subjects.
Two sizes of lateral MI were achieved by halving $r_{ab}$ and $r_{rt}$ values for the small lateral MI compared to the large one.
Two transmural extents were set by varying $r_{tm}$, which was set as 3 and 0.5 for transmural and subendocardial scars, respectively.
For baseline QRS simulation, the CV values for different directions were set as follows: 65 cm/s along the fiber direction, 48 cm/s along the sheet direction, 51 cm/s along the sheet-normal direction, and 100 cm/s and 150 cm/s for the sparse and dense endocardial directions, respectively \cite{conf/CiC/camps2022}. 
These values were consistent with reported velocities for healthy human myocardium in previous studies \cite{journal/CR/myerburg1978,journal/JMCC/taggart2000}.
In the simulation of QRS for MI, the CVs in the areas of myocardial scarring and BZ were set to 10\% and 50\% (another slower CV configuration: 5\% and 25\%) of the respective values observed in healthy myocardium.



\subsubsection{Evaluation}
For evaluation, we compared the predicted MI distribution of our proposed automatic method with the gold standard.
The ground truth of infarct regions for each scenario of each subject was obtained during simulation phase, as we explained in Sec.~\ref{method:simulation}.
To evaluate the segmentation accuracy, we calculated the Dice score, precision, and recall of the MI prediction, calculated on the PCs.
Furthermore, we propose a novel evaluation metric called the AHA-loc-score, to assess the accuracy of MI localization using the 17-segment AHA map,
\begin{equation}
	\mathrm{AHA\text{-}loc\text{-}score} = \beta_\text{c-id}\delta_\text{c-pre, c-gd} + \beta_\text{id}\text{IoU}_{id} + \beta_\text{c-d}(1-\text{d}_\text{c}),
\end{equation} 
where $\delta_\text{c-pre, c-gd}$ indicates whether the AHA index of predicted infarct center is matched with that of ground truth,
$\text{IoU}_{ids}$ calculates the intersection over union (IoU) score of the AHA indices appeared in the predicted and ground truth MI regions,
and $\text{d}_\text{c}$ refers to the normalized distance between predicted and ground truth infarct centers.
The weights $\beta_\text{c-id}$, $\beta_\text{ids}$, and $\beta_\text{c-d}$ have values of 0.5, 0.2, and 0.3, respectively.


\begin{figure*}[t] \center
    \includegraphics[width=0.99\textwidth]{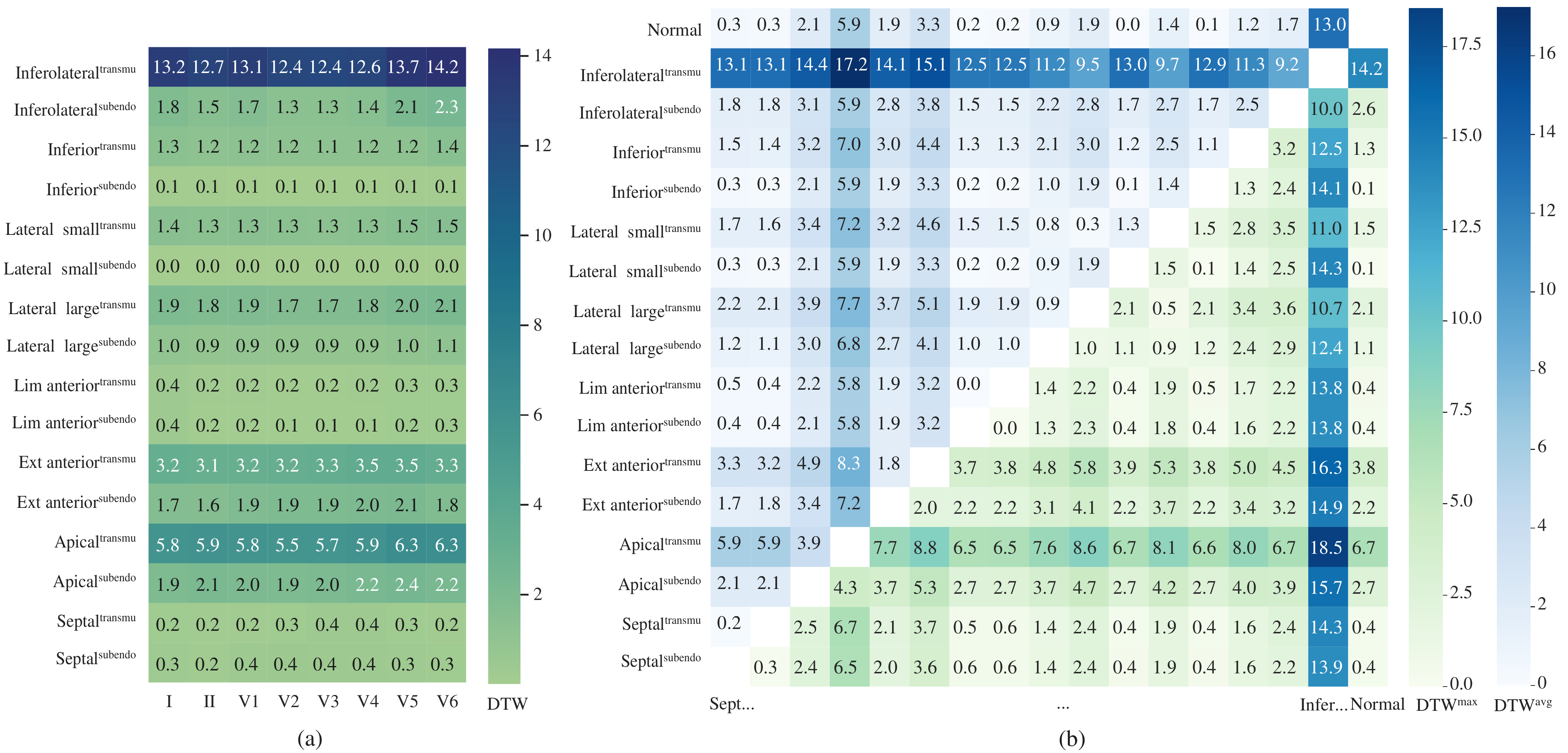}
	\caption{Comparison of QRS dissimilarity using dynamic time warping (DTW) among different MI scenarios and baseline.
    Note that here we excluded the scenario with slower conduction velocity configuration.
		(a) QRS dissimilarity of each MI scenario to the baseline in each lead;
		(b) Maximum and average QRS dissimilarity (DTW$^\text{max}$ and DTW$^\text{avg}$) between each MI scenario of all leads.  
        transmu: tranmural; subendo: subendocardial. 
		} 
	\label{fig:exp:QRS_dissimilarity}
\end{figure*}

\subsubsection{Implementation} 
The framework was implemented in PyTorch, running on a computer with 3.50~GHz Intel(R) Xeon(R) E-2146G CPU and an NVIDIA GeForce RTX 3060. 
We use the Adam optimizer to update the network parameters (weight decay = 1e-3). 
The batch size is 4, and the initial learning rate is set to 1e-4 with a stepped decay rate of 0.5 every 6800 iterations.  
The balancing parameters in Sec.~\ref{method:computational model} are set as follows: $\alpha=5$, $\lambda_\text{KL}=0.01$, $\lambda_\text{compact}=1$, $\lambda_\text{size}=1$, $\lambda_\text{spa}=1$, and $\lambda_\text{VAE}=1$.
The selection of architecture parameters was based on the property of dataset or referring previous studies or determined empirically. 
For instance, the numbers of nodes in the coarse and dense output PCs $n_\text{coarse}$ and $n_\text{dense}$ were set to 1024 and 4096, respectively, referring to Beetz et al. \cite{conf/STACOM/beetz2021}. 
The unified length of QRS signals $l_{QRS}$ was set to 512, as this was the largest length of simulated QRS signals.
The simulation of one QRS of MI spent about 5 min.
The training of the model took about 10 hours (300 epochs in total), while the inference of the networks required about 9~s to process one test case.

\subsection{Sensitivity Analysis of QRS for Different Post-MI Characteristics}  \label{method:SenAna}

We performed a sensitivity analysis in which we studied the effects of different infarct configurations in the QRS complex. 
The aim was to find out which locations and sizes had a significant effect on QRS, and thus to establish the feasibility of the inverse inference task. 
To quantify discrepancy between QRS shapes, we employed a global measure, DTW, which compared signals of different lengths with an additional penalty for the difference in QRS duration between the two signals \cite{journal/MedIA/camps2021}.
Furthermore, we introduced four QRS abnormalities reported in literature, \emph{i.e.}, \emph{QRS duration prolongation} \cite{journal/CJ/cupa2018}, \emph{pathological Q-waves} \cite{journal/JACC/delewi2013}, \emph{poor R wave progression (PRWP)} \cite{journal/IJC/kurisu2015}, and \emph{fragmented QRS (fQRS)} \cite{journal/Circulation/das2006}.

\begin{figure}[t] \center
 \includegraphics[width=0.48\textwidth]{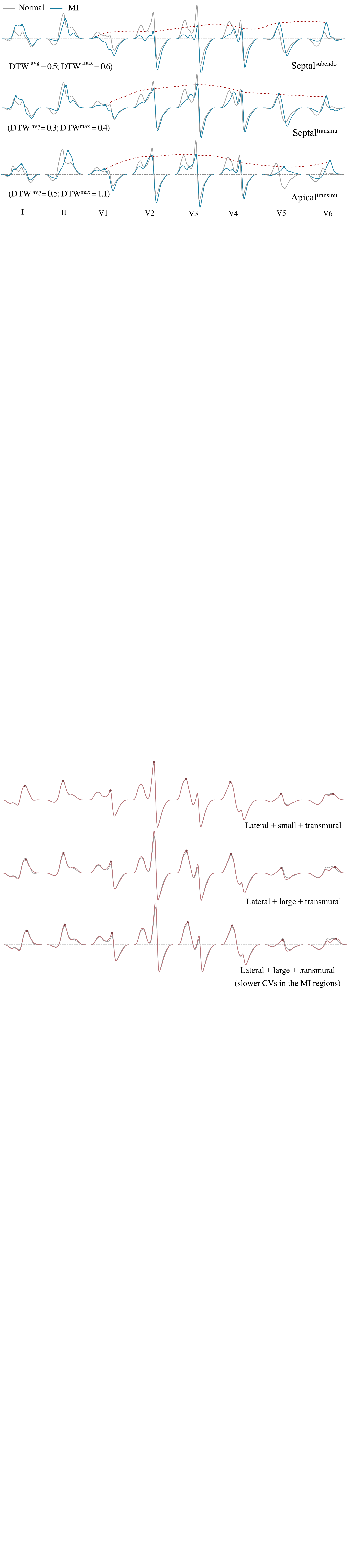}
	\caption{Comparison of QRS morphology between MI scenarios with different transmural extents and locations, along with poor R wave progression (highlighted with red dashed lines).}	
  \label{fig:exp:simulated_QRS_examples}
\end{figure}


\subsubsection{Sensitivity Analysis: Global QRS Measure}  \label{method:SenAna:global}

To assess the impact of QRS on the 17 different MI scenarios, we measured the dissimilarity between each of these and the baseline, as well as the dissimilarity between them.
As \Leireffig{fig:exp:QRS_dissimilarity} shows, the QRS complex showed morphological alterations in most post-MI scenarios when compared to the normal QRS complex (DTW $>0$). 
Particularly, inferolateral, extensive anterior, and apical transmural MI presented more evident alterations compared to others.
One can see a significant decrease in QRS morphology alteration in small lateral MI when compared to that of large lateral MI, especially for subendocardial one.
The degree of transmurality presented a noticeable impact on the QRS morphology at all infarct locations (except for limited anterior), namely transmural scars generally caused more prominent changes in QRS morphology compared to subendocardial scars.
Although the QRS dissimilarities between transmural and subendocardial septal scars were relatively small (DTW$^\text{max}=0.2$ and DTW$^\text{avg}=0.3$), differences in QRS morphology can still be observed, as shown in \Leireffig{fig:exp:simulated_QRS_examples}.
Despite the influence of transmurality on QRS morphology, the differences in QRS between various infarct locations seemed to be more pronounced than those caused by the extent of transmurality.
This implies that the QRS has greater sensitivity in localizing MI rather than predicting its transmural extent.
The primary QRS morphological difference observed with varying degrees of CV reduction was the QRS duration: 99.5 ms vs. 113.8 ms on transmural large lateral MI. 
However, our initial tests presented unexpected QRS simulation results when we significantly reduced the CVs in the MI regions. 
This suggests that the personalized CV configuration of infarct areas during simulation requires further investigation in the future.
Most infarct locations were represented on the QRS by leads I, V5, and V6, whereas septal MI was represented by leads V1-V4 and V3-V4 for subendocardial and transmural ones, respectively.
Generally, larger scars tend to result in QRS changes appearing in more leads.
This result is in agreement with those reported in clinical practice \cite{journal/CCR/nikus2014}. 

\begin{table*} [t] \center
    \caption{
    Summary of the quantitative evaluation results of our proposed method. Here, we only present the AHA-loc-score of scars.  
     }
\label{tb:results:MIinference}
{\resizebox{0.9\textwidth}{!}{	
\begin{tabular}{ll llllllll}
\hline
& \multirow{2}{*}{Scenario} & \multicolumn{3}{c}{Scar}  & \multicolumn{3}{c}{Boder zone} & \multirow{2}{*}{AHA-loc-score}\\
\hhline{~~|---| |---|~}
& & Dice & Precision & Recall & Dice & Precision & Recall & \\
\hline
\multirow{9}{*}{\rotatebox{90}{Subendocardial MI}} 
& Septal        & $ 0.349 \pm 0.198 $ & $ 0.384 \pm 0.311 $  & $ 0.517 \pm 0.327 $ & $ 0.246 \pm 0.200 $ & $ 0.236 \pm 0.237 $ & $ 0.340 \pm 0.282 $ & $ 0.687 \pm 0.347 $ \\
& Apical        & $ 0.564 \pm 0.070 $ & $ 0.490 \pm 0.154 $  & $ 0.795 \pm 0.201 $ & $ 0.336 \pm 0.064 $ & $ 0.311 \pm 0.134 $ & $ 0.497 \pm 0.177 $ & $ 0.915 \pm 0.190 $ \\
& Ext anterior  & $ 0.702 \pm 0.039 $ & $ 0.707 \pm 0.135 $  & $ 0.745 \pm 0.125 $ & $ 0.370 \pm 0.105 $ & $ 0.391 \pm 0.102 $ & $ 0.389 \pm 0.129 $ & $ 0.721 \pm 0.264 $ \\
& Lim anterior  & $ 0.476 \pm 0.241 $ & $ 0.464 \pm 0.201 $  & $ 0.584 \pm 0.349 $ & $ 0.284 \pm 0.122 $ & $ 0.266 \pm 0.157 $ & $ 0.272 \pm 0.173 $ & $ 0.691 \pm 0.328 $ \\
& Lateral large & $ 0.222 \pm 0.219 $ & $ 0.200 \pm 0.149 $  & $ 0.323 \pm 0.370 $ & $ 0.137 \pm 0.138 $ & $ 0.147 \pm 0.159 $ & $ 0.206 \pm 0.265 $ & $ 0.452 \pm 0.288 $ \\
& Lateral small & $ 0.097 \pm 0.112 $ & $ 0.067 \pm 0.087 $  & $ 0.388 \pm 0.452 $ & $ 0.000 \pm 0.000 $ & $ 0.000 \pm 0.000 $ & $ 0.000 \pm 0.000 $ & $ 0.384 \pm 0.286 $ \\
& Inferior      & $ 0.228 \pm 0.252 $ & $ 0.233 \pm 0.301 $  & $ 0.295 \pm 0.323 $ & $ 0.100 \pm 0.144 $ & $ 0.130 \pm 0.212 $ & $ 0.120 \pm 0.159 $ & $ 0.480 \pm 0.361 $ \\
& Inferolateral & $ 0.527 \pm 0.256 $ & $ 0.539 \pm 0.283 $  & $ 0.602 \pm 0.330 $ & $ 0.321 \pm 0.158 $ & $ 0.346 \pm 0.184 $ & $ 0.370 \pm 0.196 $ & $ 0.550 \pm 0.316 $ \\
\cline{2-9}
& \textit{Average} & $ 0.396 \pm 0.271 $ & $ 0.386 \pm 0.292 $  & $ 0.531 \pm 0.368 $ & $ 0.220 \pm 0.178 $ & $ 0.228 \pm 0.203 $ & $ 0.274 \pm 0.243 $ & $ 0.610 \pm 0.343 $ \\
\hline 
\multirow{9}{*}{\rotatebox{90}{Transmural MI}} 
& Septal        & $ 0.739 \pm 0.217 $ & $ 0.814 \pm 0.195 $  & $ 0.695 \pm 0.195 $ & $ 0.698 \pm 0.241 $ & $ 0.628 \pm 0.244 $ & $ 0.628 \pm 0.244 $ & $ 0.680 \pm 0.281 $ \\
& Apical        & $ 0.779 \pm 0.186 $ & $ 0.950 \pm 0.046 $  & $ 0.703 \pm 0.221 $ & $ 0.699 \pm 0.140 $ & $ 0.796 \pm 0.077 $ & $ 0.649 \pm 0.187 $ & $ 0.921 \pm 0.179 $ \\
& Ext anterior  & $ 0.934 \pm 0.028 $ & $ 0.967 \pm 0.020 $  & $ 0.908 \pm 0.057 $ & $ 0.785 \pm 0.043 $ & $ 0.761 \pm 0.041 $ & $ 0.815 \pm 0.061 $ & $ 0.987 \pm 0.007 $ \\
& Lim anterior  & $ 0.450 \pm 0.280 $ & $ 0.696 \pm 0.412 $  & $ 0.392 \pm 0.274 $ & $ 0.242 \pm 0.162 $ & $ 0.430 \pm 0.259 $ & $ 0.193 \pm 0.137 $ & $ 0.711 \pm 0.333 $ \\
& Lateral large & $ 0.211 \pm 0.220 $ & $ 0.397 \pm 0.381 $  & $ 0.168 \pm 0.185 $ & $ 0.132 \pm 0.129 $ & $ 0.437 \pm 0.340 $ & $ 0.091 \pm 0.107 $ & $ 0.403 \pm 0.247 $ \\
& Lateral small & $ 0.311 \pm 0.244 $ & $ 0.363 \pm 0.289 $  & $ 0.372 \pm 0.353 $ & $ 0.011 \pm 0.032 $ & $ 0.025 \pm 0.075 $ & $ 0.010 \pm 0.029 $ & $ 0.609 \pm 0.328 $ \\
& Inferior      & $ 0.173 \pm 0.288 $ & $ 0.162 \pm 0.263 $  & $ 0.195 \pm 0.332 $ & $ 0.106 \pm 0.193 $ & $ 0.149 \pm 0.271 $ & $ 0.111 \pm 0.207 $ & $ 0.270 \pm 0.227 $ \\
& Inferolateral & $ 0.543 \pm 0.245 $ & $ 0.874 \pm 0.247 $  & $ 0.437 \pm 0.275 $ & $ 0.452 \pm 0.188 $ & $ 0.689 \pm 0.182 $ & $ 0.374 \pm 0.234 $ & $ 0.688 \pm 0.284 $ \\
\cline{2-9}
& \textit{Average} & $ 0.518 \pm 0.347 $ & $ 0.653 \pm 0.391 $ & $ 0.484 \pm 0.354 $ & $ 0.385 \pm 0.322 $ & $ 0.498 \pm 0.340 $ & $ 0.359 \pm 0.331 $ & $ 0.659 \pm 0.339 $ \\
\hline \hline
& \textbf{Average} & $ 0.457 \pm 0.317 $ & $ 0.519 \pm 0.370 $ & $ 0.507 \pm 0.362 $ & $ 0.302 \pm 0.273 $ & $ 0.363 \pm 0.311 $ & $ 0.317 \pm 0.293 $ & $ 0.634 \pm 0.342 $ \\
\hline
\end{tabular} }
}
\end{table*}

\subsubsection{Sensitivity Analysis: Local QRS Measure}  \label{method:SenAna:local}

The changes in QRS morphology for the 17 MI scenarios were reflected in multiple ways.
Here, we introduced several QRS criteria and compared the contribution of each of these for infarct detection.
We found that apical and inferolateral MI tended to present prolongation of the QRS duration: 124.1 ms and 107.7 ms (apical and inferolateral MI) vs. 90.4 ms (normal).
PRWP mainly occurred in extensive anterior, septal, and apical MI, similar as reported in the literature \cite{journal/IJC/kurisu2015,journal/IJC/mittal1986}. 
Specifically, the R wave amplitude in the septal MI was sometimes flattened, while the R wave of V6 tended to be larger than that of V5 in the apical MI, as \Leireffig{fig:exp:simulated_QRS_examples} shows.
The prevalence of fQRS was more common in the inferior lead (lead II) compared with the anterior leads (leads V3 and V4) and the lateral leads (leads V5 and V6), similar to the results reported in Liu \textit{et al.} \cite{journal/FiP/luo2020}.
The presence of fQRS in lead II and leads V3-V4 indicated inferolateral and extensive anterior MI, respectively.
In contrast, pathological Q wave failed to classify MI from healthy subjects in our simulation system.

\begin{figure*}[!t]\center
\includegraphics[width=0.78\textwidth]{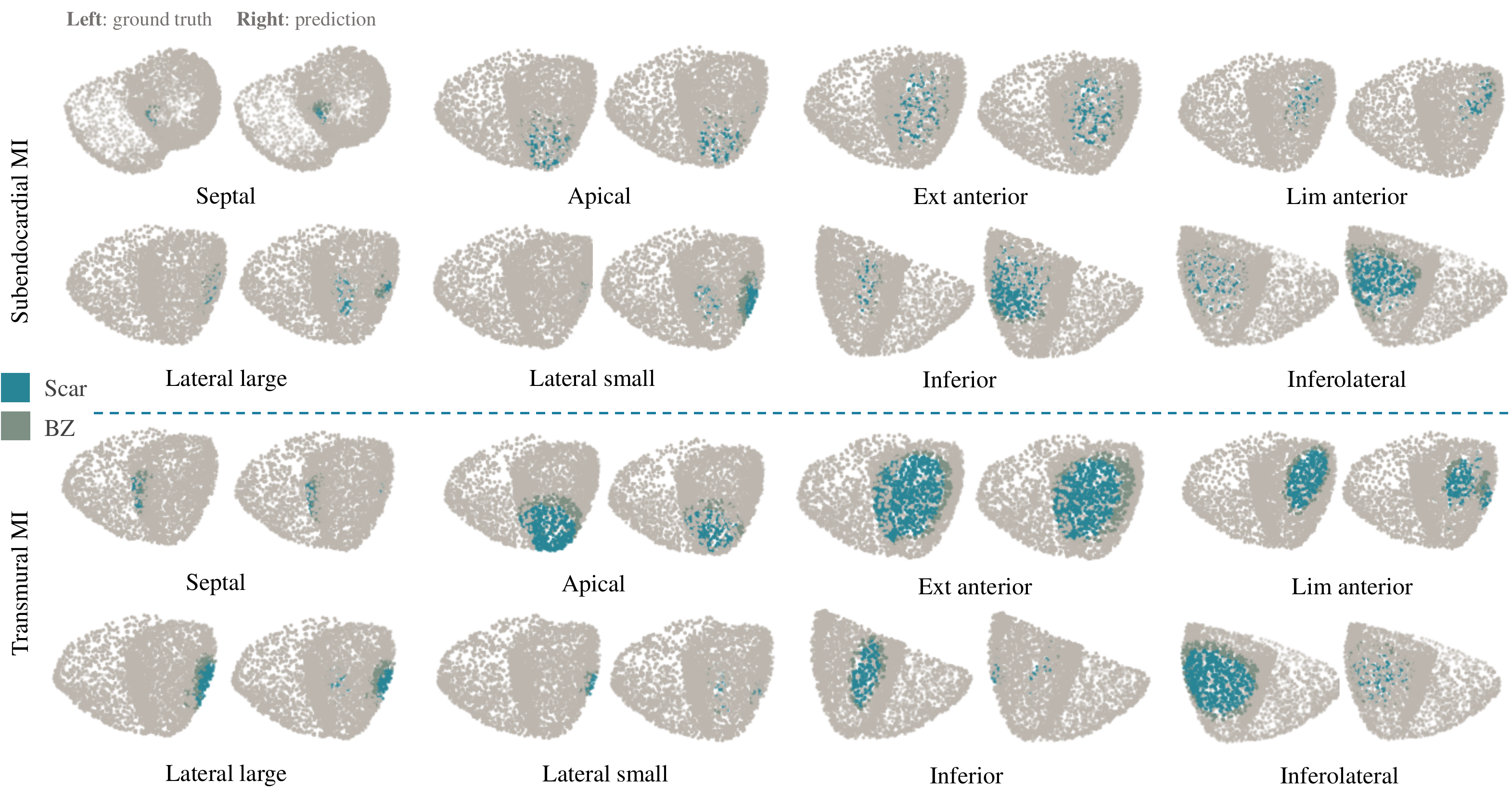}
   \caption{3D visualization of infarct detection results using the proposed method. For each scenario, the left side is the ground truth, and the right side is the prediction.}
	\label{fig:result:3d_visualization_MI}
\end{figure*}

\subsection{Inference Accuracy of Post-MI Properties} 

\begin{figure*}[t]\center
    \includegraphics[width=0.98\textwidth]{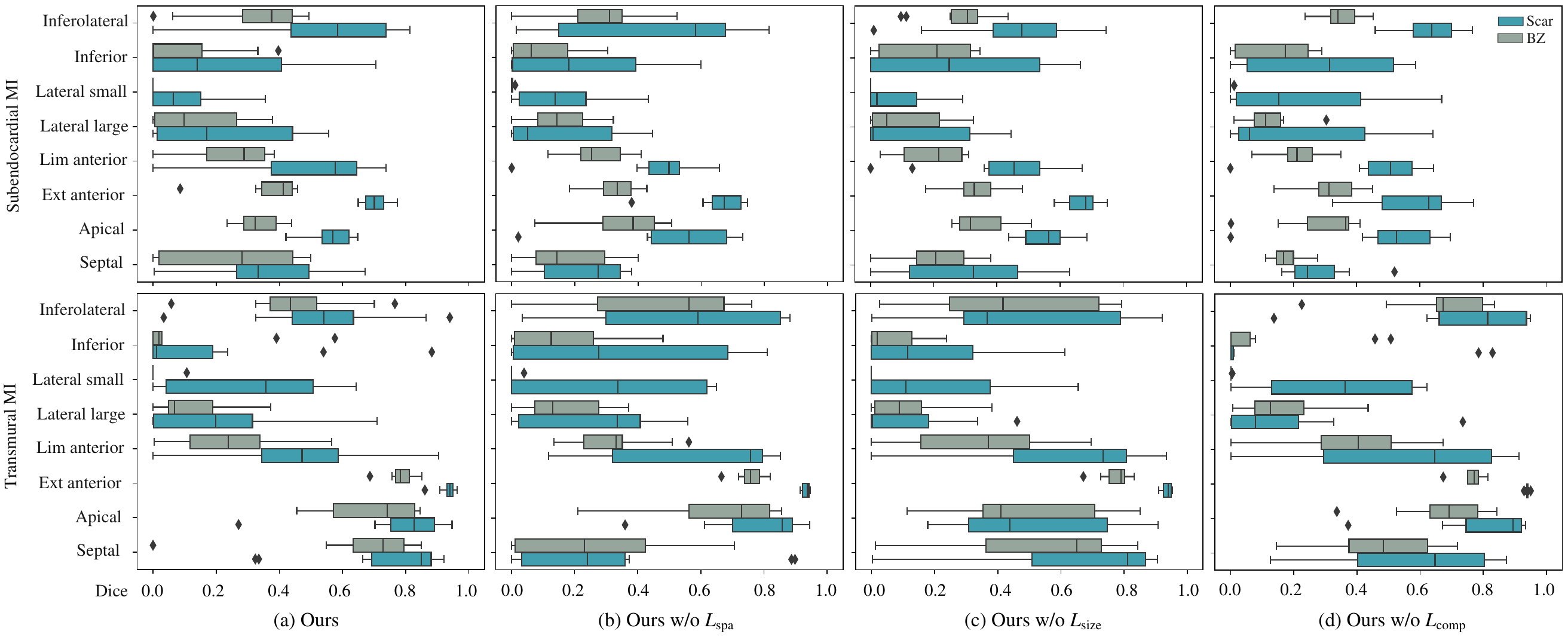}
   \caption{
   Comparison of boxplots between the proposed method and the methods in the ablation study in terms of Dice scores.
   }
\label{fig:result:ablation_study}
\end{figure*}

\Leireftb{tb:results:MIinference} presents the quantitative results of the proposed method, and \Leireffig{fig:result:ablation_study} (a) provides the boxplots of Dice score.
The proposed method obtained the best segmentation and localization performance on the transmural extensive anterior MI (Dice= $ 0.934 \pm 0.028 $, AHA-loc-score~$= 0.987 \pm 0.007$).
Even for the scenarios where there were not notable QRS morphology changes, such as MI in the septum and limited anterior areas, the model still can localize the corresponding infarct (DTW$^\text{max}=0.4$, AHA-loc-score~$ \approx 0.7$).
Nevertheless, the model showed difficulties in detecting lateral (especially for the subendocardial and small size ones, with Dice score of $ 0.097 \pm 0.112 $) and inferior MI with Dice scores of $ 0.228 \pm 0.252 $ and $ 0.173 \pm 0.288 $ for subendocardial and transmural one, respectively.
In general, the segmentation of the transmural MI tended to be more accurate than that of the subendocardial MI (Dice: $ 0.518 \pm 0.347 $ vs. $ 0.396 \pm 0.271 $). 
This observation aligned with expectations, since transmural MI often exhibit more pronounced and distinct QRS abnormalities compared to subendocardial MI, as proved in previous sensitivity analysis. 
As a result, our model can leverage these noticeable differences to identify and segment the affected region accurately.
Nevertheless, their ability to precisely determine the location of the infarction within the myocardium did not vary significantly (AHA-loc score: $ 0.610 \pm 0.343 $ vs. $ 0.659 \pm 0.339 $).
This can be attributed to the fact that the localization of MI is not solely dependent on the depth or extent of the infarct.
Furthermore, the accuracy of predicting scars was generally higher than that of predicting border zones (BZs).
This could be because the complex nature of BZs, where the myocardial tissue undergoes a transition from healthy to scarred, introduces additional variability and ambiguity in the QRS signals, leading to a lower prediction accuracy for BZs.
The performance in terms of Dice coefficient, precision, recall and AHA-loc-score was generally consistent.
However, in specific cases like apical, limited anterior, and inferolateral transmural MI, precision may exhibit a slight superiority over the Dice. 
Apical MI obtained the highest AHA-loc-score, indicating its accurate and reliable localization. 
This could be attributed to the uniqueness of the apical location, allowing for a more precise and unambiguous localization of MI due to the absence of significant interference from neighboring structures.

Figure~\ref{fig:result:3d_visualization_MI} provides 3D  results of a representative test subject on different scenarios.
One can observe that the 3D visualization agrees well with the quantitative analysis result.
There were outliers appearing in the inferior area for lateral MI detection and vice versa, which suggests that the model had difficulty distinguishing between the lateral and inferior MI areas based on their QRS.
Furthermore, even though extensive anterior and inferolateral MI both covered large areas, the detection of inferolateral MI tended to be more difficult compared to that of extensive anterior MI.

\subsection{Ablation Study} 
Accurate MI inference goes beyond merely identifying the location of the infarction, but also requires a comprehensive assessment of the extent of infarct tissue.
Therefore, we introduced additional constrains, namely localization constrains ($\mathcal{L}_\text{spa}$ and $\mathcal{L}_\text{comp}$) and an extent constrain ($\mathcal{L}_\text{size}$).
To evaluate their effectiveness, we conducted an ablation study by selectively removing them from the proposed framework, as presented in \Leireffig{fig:result:ablation_study}.
One can see that in most scenarios the proposed method obtained the best performance compared to others.
For example, without localization constrains, the model presented worse performance in identifying septal MI.
Note that septal MI normally presents complexity for detection, due to its unique position and overlapping ECG effects from neighboring regions, such as the anterior and inferior walls \cite{conf/MICCAI/xu2014,journal/TMI/ghimire2019}.
We observed that the absence of $\mathcal{L}_\text{comp}$ led to improved Dice in cases of inferolateral and subendocardial limited anterior MI and decreased Dice in cases of extensive anterior MI.
Nevertheless, reduction in outliers observed in the visualization results suggests that $\mathcal{L}_\text{comp}$ effectively minimizes the occurrence of outliers, leading to more reliable and accurate predictions.
The extent constraint was also crucial, particularly in distinguishing between subendocardial and transmural MI that present different sizes in the same anatomical position.

\begin{figure*}[t]\center
    \subfigure[] {\includegraphics[width=0.46\textwidth]{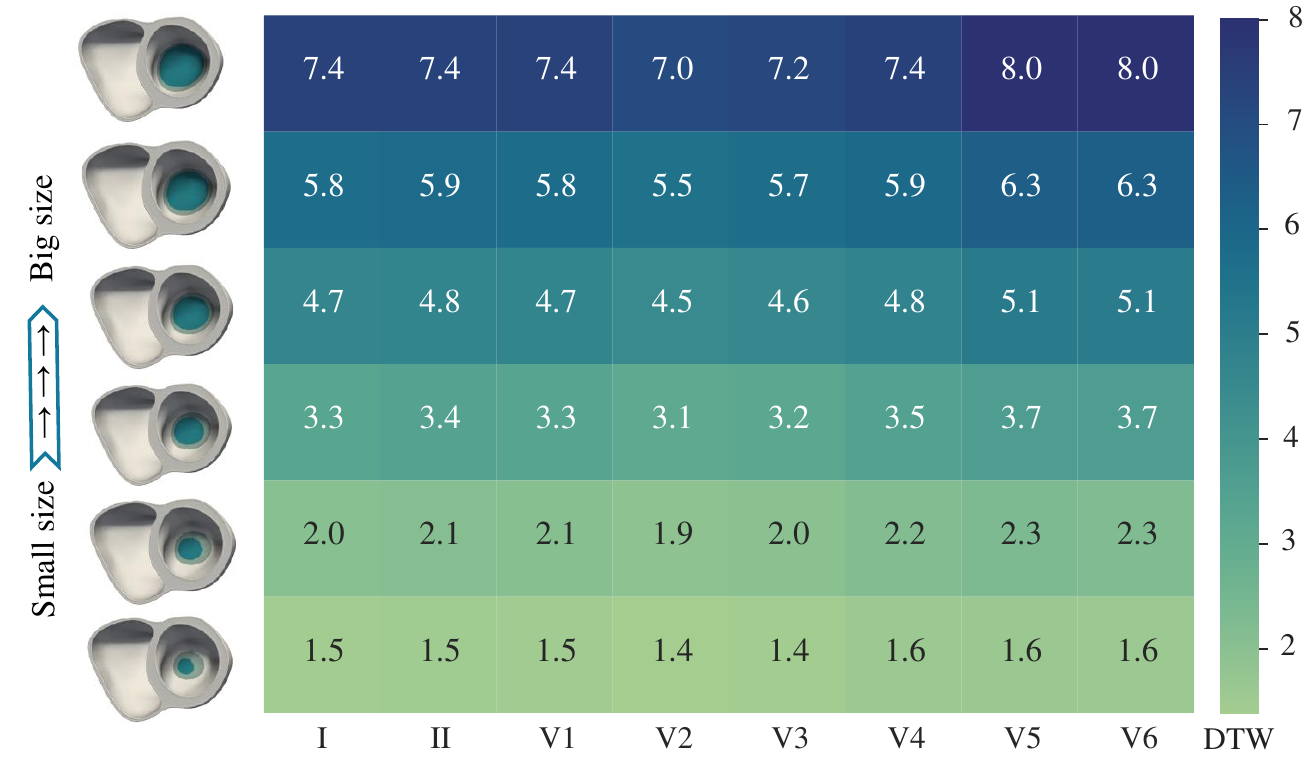}}
    \subfigure[] {\includegraphics[width=0.37\textwidth]{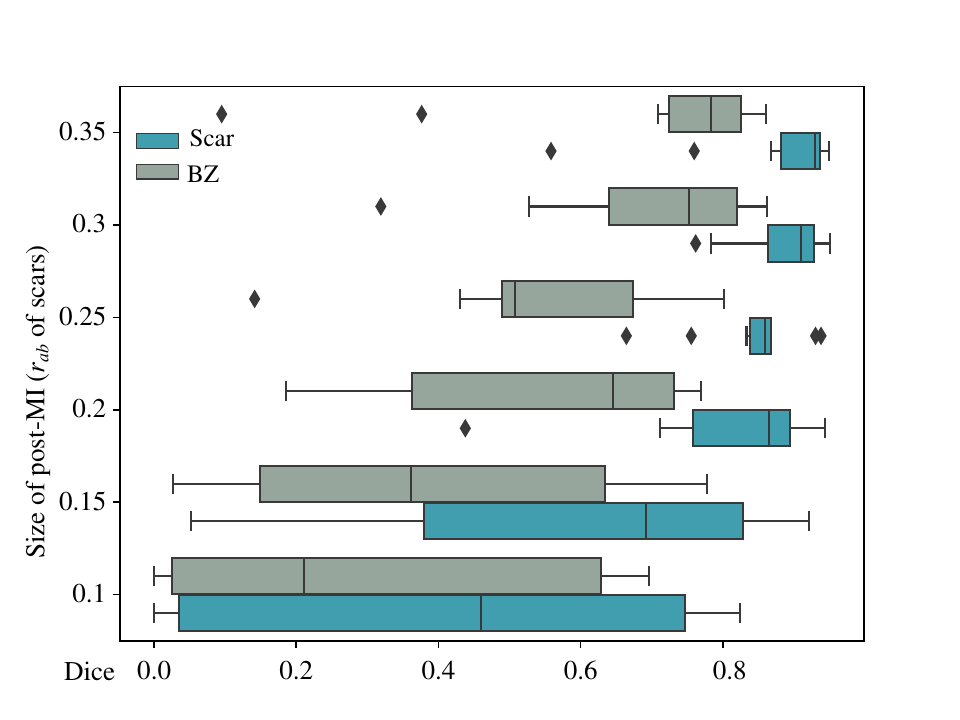}}
    \subfigure[] {\includegraphics[width=1\textwidth]{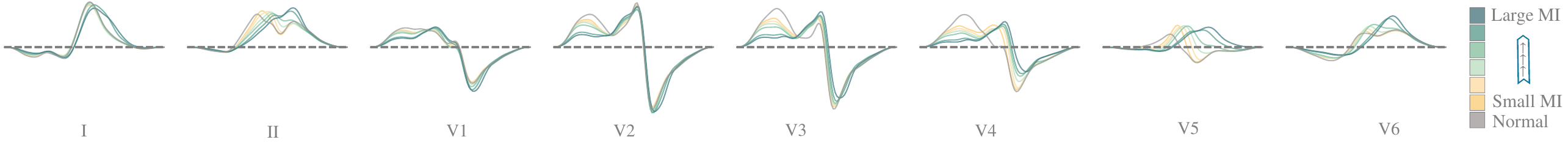}}
   \caption{
   The effect of infarct size on the simulated QRS morphology and the inference accuracy of MI.
   (a) Illustration of the different infarct sizes and their corresponding QRS morphological dissimilarities to the baseline in each lead. 
   (b) Boxplot of Dice scores of the proposed method for the inference of scars/ BZ with different sizes.
   (c) Comparison of QRS morphology of baseline and MI with different infarct sizes. 
   }
\label{fig:result:QRS_MIsize}
\end{figure*}

\subsection{Extended Evaluation}




\subsubsection{Exploring the Detection Limit of QRS for Small Infarct Areas}
To investigate what is the smallest infarct area that can be detected from QRS complexes, we employed apical MI as an example and varied the infarct size and retrained the model based on the pre-trained one. 
The idea behind this approach is to determine the sensitivity of QRS-based detection methods for small infarct areas, which may have important clinical implications for risk stratification and management of post-MI patients.
Figures~\ref{fig:result:QRS_MIsize} (a) and (c) demonstrate that as the infarct size decreased, the QRS morphological changes also diminished.
This is because a smaller infarct would have a lesser impact on the overall electrical conduction and activation patterns of the heart. 
Consequently, the deviations in the QRS, which represent the depolarization of the ventricles, would be less pronounced. 
Nevertheless, our method still can extract subtle features from the QRS complex that may be indicative of small infarct areas, as \Leireffig{fig:result:QRS_MIsize} (b) shows.
This ability was limited until when the Cobiveco apicobasal radius $r_{ab}$ of scars equaled to 0.1 for apical MI.

\subsubsection{Correlation Analysis: Relationship between QRS/ PC Reconstruction and MI Inference Accuracy}
To evaluate the robustness of the proposed inference scheme to the reconstruction error, we analyzed the relationship between the reconstruction and inference errors by the proposed method. 
The accuracy of PC and QRS reconstruction was calculated as 0.5*$\mathcal{L}^\text{rec}_\text{PC}$ with $\alpha=1$ and $\mathcal{L}^\text{rec}_\text{QRS}$, respectively.
The $r^2$ values of scar/ BZ for PC and QRS-MI inference correlations were 0.002/ 0.006 and 0.008/ 0.009, respectively, indicating no relationship between inference and reconstruction accuracy. 
This implies that the accuracy of MI inference using the proposed method was not significantly influenced by the quality of the reconstruction.
This is reasonable, as the proposed method focuses on extracting relevant features from the input data rather than relying solely on accurate reconstruction for MI inference.
Nevertheless, the reconstructions are still necessary to embed both anatomical and electrophysiological features into the unified latent space for the inference. 
To demonstrate this, we removed one of the reconstructions, such as PC reconstruction, for comparison.
By solely embedding QRS features into the latent space, we found that the QRS reconstruction accuracy was not statistically different ($ 1.427 \pm 0.657 $ vs. $ 1.420 \pm 0.540 $, p-value = 0.916), while the performance of MI inverse inference was negatively affected (AHA-loc scores: $ 0.634 \pm 0.342 $ vs. $ 0.552 \pm 0.335 $).

\begin{table*} [t] \center
    \caption{
    Overview of previous methods for MI prediction. IHD: ischemic heart disease; T2W: T2-weighted; $^\dagger$: semi-supervised.} 
\label{tb:discussion:review}
{\small
\begin{tabular}{ l l *{3}{@{\ \,} l }}\hline
Study     &  No. subjects &  Input modality & Result (Dice) \\
\hline
Ukwatta \textit{et al.} (2015)$^\dagger$ \cite{journal/TMI/ukwatta2015} & 61 IHD and LV dysfunction             & LGE MRI            & Scar: $ 0.880 \pm 0.077 $; BZ: $ 0.653 \pm 0.085 $ \\ 
Rajchl \textit{et al.} (2013) \cite{journal/TMI/rajchl2013}             & 35 MI + 15 Tetralogy of Fallot        & LGE MRI            & Scar: $ 0.776 \pm 0.033 $ \\
Moccia \textit{et al.} (2019) \cite{journal/MRMPBM/moccia2019}          & 30 IHD                                & LGE MRI            & Scar: $ 0.712 $ \\
Zhang \textit{et al.} (2019) \cite{journal/Radiology/zhang2019}         & 212 chronic MI + 87 normal            & Cine MRI           & Scar: $ 0.861 \pm 0.057 $ \\ 
Xu \textit{et al.} (2020) \cite{journal/MedIA/xu2020a}                  & 230 IHD + 50 normal                   & Cine MRI           & Scar: $ 0.932 \pm 0.110 $ \\ 
Popescu \textit{et al.} (2022) \cite{journal/CDHJ/popescu2022}          & 155 IHD                               & Cine, LGE MRI      & Scar: $ 0.570 \pm 0.050 $ \\ 
Wang \textit{et al.} (2022) \cite{journal/MedIA/wang2022}               & 45 acute MI                           & Cine, LGE, T2W MRI & Scar: $ 0.678 \pm 0.242 $;  Scar \& BZ: $ 0.735 \pm 0.111 $ \\ 
Ding \textit{et al.} (2023) \cite{journal/TMI/ding2023}                 & 45 (+50) acute MI                     & Cine, LGE, T2W MRI & Scar: $ 0.649 \pm 0.098 $; Scar \& BZ: $ 0.760 \pm 0.098 $ \\
Ghimire \textit{et al.} (2019) \cite{journal/TMI/ghimire2019}           & Synthetic data                        & CT, 120-lead ECG   & Scar: $ 0.42 \pm 0.16 $ \\
\hline
\end{tabular} }
\end{table*}

\section{Discussion and Conclusion}
In this paper, we have developed a deep computational model to tackle the inverse problem in cardiac electrophysiology, \emph{i.e.}, inferring MI distribution from QRS signals.
Through the integration of anatomical and electrophysiological data, we achieve a comprehensive analysis that incorporates different infarct locations, sizes, transmural extents, and cardiac electrical activity alterations. 
By consistently representing the ventricular anatomy in a coordinate reference system, we establish a robust sensitivity analysis framework for studying the association between infarct characteristics and QRS abnormalities.
The sensitivity analysis results have demonstrated significant morphological alterations in the QRS complex for various post-MI scenarios, particularly inferolateral, extensive anterior, and apical MI. 
These findings suggest that the involvement of large areas of damaged heart muscle leads to pronounced changes in QRS morphology. 
Furthermore, the analysis emphasizes the impact of transmurality on QRS morphology, namely transmural MI presents more prominent changes compared to subendocardial MI.
However, the differences in QRS between various infarct locations can be more pronounced than those caused by the extent of transmurality, indicating the greater sensitivity of QRS in localizing MI rather than predicting its transmural extent.
The analysis further highlights the importance of lead selection in accurately detecting the location of infarction. 
Overall, the sensitivity analysis provides valuable insights into the relationship between infarct characteristics and QRS abnormalities, enhancing our understanding of the complex interplay between infarct characteristics and electrophysiological features. 

The proposed method can effectively segment and localize MI, even in scenarios with limited QRS morphology changes, demonstrating its feasibility of developing CDTs for MI patients.
The results of the ablation study emphasize the importance of the localization and extent constraints in accurate MI inference. 
The proposed method exhibits the ability to detect small infarct areas, although its sensitivity is limited, as proved in our extended study. 
The correlation analysis demonstrates that while incorporating reconstruction in the inference process is important, the accuracy of MI inference is not significantly dependent on the quality of reconstruction.
To conduct a sensitivity analysis of MI properties, we intentionally select consistent infarct location, size and transmural extent for each subject.
While it ensures a controlled comparison, it may have led to a limited evaluation of MI inference.
We conduct a small test by randomly selecting infarct for each subject and only obtain reasonable good results on few cases.
This outcome is expected because randomly simulating a single scenario for each subject limits ability of the proposed model to learn and generalize across different infarct characteristics. 
In order to improve performance, in the future a more diverse and comprehensive dataset with a wider range of infarct scenarios should be used to train the model.

\Leireftb{tb:discussion:review} summarizes the related works from literature.
Ukwatta \textit{et al.} \cite{journal/TMI/ukwatta2015} evaluated their method on a dataset consisting of 61 LGE MRIs and obtained average Dice scores of $ 0.880 \pm 0.077 $ and $ 0.653 \pm 0.085 $ for scar and BZ segmentation, respectively.
Their method required an accurate initialization of LV myocardium from manual segmentation, followed by a continuous max-flow based classification.
In contrast, the other two fully supervised LGE MRI based LV scar segmentation model reported comparatively lower Dice scores: $ 0.776 \pm 0.033 $ \cite{journal/TMI/rajchl2013} and $ 0.712 $ \cite{journal/MRMPBM/moccia2019}.
Alternatively, Zhang \textit{et al.} \cite{journal/Radiology/zhang2019} and Xu \textit{et al.} \cite{journal/MedIA/xu2020a} utilized cine MRI to segment LV scars and obtained quite promising results in terms of Dice score, i.e., $ 0.861 \pm 0.057 $ and $ 0.932 \pm 0.110 $, respectively.
Popescu \textit{et al.} \cite{journal/CDHJ/popescu2022} performed style transfer to generate synthetic LGE images from cine MRI for LGE data augmentation, and they reported a Dice score $ 0.570 \pm 0.050 $.
Furthermore, Wang \textit{et al.} \cite{journal/MedIA/wang2022} and Ding \textit{et al.} \cite{journal/TMI/ding2023} combine the complementary information from multi-sequence MRI for the myocardial pathology segmentation.
They reported $ 0.678 \pm 0.242 $/ $ 0.735 \pm 0.111 $ and $ 0.649 \pm 0.098 $/ $ 0.760 \pm 0.098 $ Dice scores for LV scar/ scar+BZ segmentation, respectively.
It should be noted that all these eight works segment scar/ BZ from images which usually include limited slices, and only Ukwatta \textit{et al.} \cite{journal/TMI/ukwatta2015} reconstructed 3D geometry of infarct from its segmentation.
Ghimire \textit{et al.} \cite{journal/TMI/ghimire2019} segmented the infact area on a image-driven 3D heart-torso models from simulated 120-lead ECG, and reported a Dice score of scars $ 0.42 \pm 0.16 $.
We also computed the average Dice scores (Scar/ BZ: $ 0.457 \pm 0.317 $/ $ 0.302 \pm 0.273 $) for infarct segmentation based on 3D geometry, which can be directly integrated into personalized cardiac electrophysiology modeling.
Note that it can be difficult to pursue an objective cross-study comparison due to the difference of datasets, initialization methods, and definition of evaluation metrics.

This study represents a proof-of-concept investigation, and it is essential to acknowledge several inherent limitations.
Firstly, this study assumes a known set of RNs and fixed CVs for all subjects, which may not fully capture the complexity and heterogeneity present in real-world healthcare data. 
Therefore, further research is needed to personalize these activation properties based on individual patient characteristics and specific healthcare settings. 
Secondly, we only consider cardiac anatomical information and electrode nodes while disregarding the full torso geometry.
The inclusion of torso geometry could provide valuable insights into its influence on QRS patterns.
By incorporating full torso geometry in our future work, we can gain a more comprehensive understanding of the factors influencing QRS patterns and improve the accuracy of our predictions and interpretations.
Moreover, we only reconstructed cardiac anatomy in the ED phase for the ECG simulation and inverse inference, as the mesh reconstruction from the MRI is time-expensive.
In the future, we can introduce efficient patient-specific 4D (3D + time) cardiac mesh reconstruction algorithms for mechanistic modelling, which describes the contraction and relaxation of the myocardium.
Thirdly, this study focuses solely on the QRS complex, rather than considering the entire ECG signal.
Applying the analysis to the whole ECG signal would provide a more comprehensive assessment but may require significant computational resources.
To address this limitation, future research could explore computationally efficient surrogate to replace the expensive simulation model.
Fourthly, the straightforward concatenation of the acquired features from PC and QRS lacks a comprehensive explanation regarding the alignment and holistic contribution of these features to the MI inference. 
Therefore, in the future we will extend current study by integrating domain-specific knowledge and principles derived from the underlying physics of the cardiac system. 
By explicitly encoding the physical interactions and dependencies between QRS and anatomy, we aim to establish a more transparent and interpretable way for capturing the intricate interplay between the two modalities for the inverse inference.
Finally, while the developed CDTs can provide valuable insights into the mechanisms of MI, they are based on simplified assumptions about the heart and may not capture all aspects of the complex interactions between cardiac structures and functions.
Given the limitations, particularly in the simulated dataset used, this can only serve as a proof of concept until validation on the clinical data can be performed.
    
\bibliographystyle{ieeetr}
\bibliography{tmi_refs}

\begin{thebibliography}{10}

\bibitem{journal/lancet/john2012}
R.~M. John, U.~B. Tedrow, B.~A. Koplan, {\em et~al.}, ``Ventricular arrhythmias and sudden cardiac death,'' {\em The Lancet}, vol.~380, no.~9852, pp.~1520--1529, 2012.

\bibitem{journal/MedIA/li2023}
L.~Li, F.~Wu, S.~Wang, {\em et~al.}, ``Myo{PS}: A benchmark of myocardial pathology segmentation combining three-sequence cardiac magnetic resonance images,'' {\em Medical Image Analysis}, p.~102808, 2023.

\bibitem{journal/Radiology/ordovas2011}
K.~G. Ordovas and C.~B. Higgins, ``Delayed contrast enhancement on mr images of myocardium: past, present, future,'' {\em Radiology}, vol.~261, no.~2, pp.~358--374, 2011.

\bibitem{journal/Radiology/zhang2019}
N.~Zhang, G.~Yang, Z.~Gao, {\em et~al.}, ``Deep learning for diagnosis of chronic myocardial infarction on nonenhanced cardiac cine {MRI},'' {\em Radiology}, vol.~291, no.~3, pp.~606--617, 2019.

\bibitem{journal/MedIA/xu2020a}
C.~Xu, L.~Xu, {\em et~al.}, ``Contrast agent-free synthesis and segmentation of ischemic heart disease images using progressive sequential causal {GAN}s,'' {\em Medical Image Analysis}, vol.~62, p.~101668, 2020.

\bibitem{journal/NEJM/zimetbaum2003}
P.~J. Zimetbaum and M.~E. Josephson, ``Use of the electrocardiogram in acute myocardial infarction,'' {\em New England Journal of Medicine}, vol.~348, no.~10, pp.~933--940, 2003.

\bibitem{journal/JE/hanna2011}
E.~B. Hanna and D.~L. Glancy, ``{ST}-segment depression and {T}-wave inversion: classification, differential diagnosis, and caveats,'' {\em Cleveland Clinic Journal of Medicine}, vol.~78, no.~6, p.~404, 2011.

\bibitem{conf/FIMH/li2023}
L.~Li, J.~Camps, {\em et~al.}, ``Influence of myocardial infarction on {QRS} properties: A simulation study,'' in {\em International Conference on Functional Imaging and Modeling of the Heart}, pp.~223--232, 2023.

\bibitem{journal/EHJ/corral2020}
J.~Corral-Acero, F.~Margara, M.~Marciniak, {\em et~al.}, ``The ‘digital twin’ to enable the vision of precision cardiology,'' {\em European Heart Journal}, vol.~41, no.~48, pp.~4556--4564, 2020.

\bibitem{journal/NC/arevalo2016}
H.~J. Arevalo, F.~Vadakkumpadan, {\em et~al.}, ``Arrhythmia risk stratification of patients after myocardial infarction using personalized heart models,'' {\em Nature Communications}, vol.~7, no.~1, p.~11437, 2016.

\bibitem{journal/MedIA/gillette2021}
K.~Gillette, M.~A. Gsell, A.~J. Prassl, {\em et~al.}, ``A framework for the generation of digital twins of cardiac electrophysiology from clinical 12-leads {ECG}s,'' {\em Medical Image Analysis}, vol.~71, p.~102080, 2021.

\bibitem{journal/MedIA/qiu2023}
J.~Qiu, L.~Li, {\em et~al.}, ``Myo{PS-Net}: Myocardial pathology segmentation with flexible combination of multi-sequence cmr images,'' {\em Medical Image Analysis}, vol.~84, p.~102694, 2023.

\bibitem{journal/MedIA/wang2022}
K.-N. Wang, X.~Yang, {\em et~al.}, ``A{WS}net: an auto-weighted supervision attention network for myocardial scar and edema segmentation in multi-sequence cardiac magnetic resonance images,'' {\em Medical Image Analysis}, vol.~77, p.~102362, 2022.

\bibitem{journal/TMI/ding2023}
W.~Ding, L.~Li, {\em et~al.}, ``Aligning multi-sequence {CMR} towards fully automated myocardial pathology segmentation,'' {\em IEEE Transactions on Medical Imaging}, 2023.

\bibitem{journal/MedIA/karim2016}
R.~Karim, P.~Bhagirath, P.~Claus, {\em et~al.}, ``Evaluation of state-of-the-art segmentation algorithms for left ventricle infarct from late gadolinium enhancement {MR} images,'' {\em Medical image analysis}, vol.~30, pp.~95--107, 2016.

\bibitem{journal/MRM/tao2010}
Q.~Tao, J.~Milles, K.~Zeppenfeld, {\em et~al.}, ``Automated segmentation of myocardial scar in late enhancement {MRI} using combined intensity and spatial information,'' {\em Magnetic Resonance in Medicine}, vol.~64, no.~2, pp.~586--594, 2010.

\bibitem{conf/CC/baron2008}
N.~Baron, N.~Kachenoura, F.~Beygui, {\em et~al.}, ``Quantification of myocardial edema and necrosis during acute myocardial infarction,'' in {\em 2008 Computers in Cardiology}, pp.~781--784, IEEE, 2008.

\bibitem{journal/TMI/ukwatta2015}
E.~Ukwatta, H.~Arevalo, {\em et~al.}, ``Myocardial infarct segmentation from magnetic resonance images for personalized modeling of cardiac electrophysiology,'' {\em IEEE transactions on medical imaging}, vol.~35, no.~6, pp.~1408--1419, 2015.

\bibitem{journal/TMI/rajchl2013}
M.~Rajchl, J.~Yuan, {\em et~al.}, ``Interactive hierarchical-flow segmentation of scar tissue from late-enhancement cardiac {MR} images,'' {\em IEEE transactions on medical imaging}, vol.~33, no.~1, pp.~159--172, 2013.

\bibitem{journal/MRMPBM/moccia2019}
S.~Moccia, R.~Banali, {\em et~al.}, ``Development and testing of a deep learning-based strategy for scar segmentation on {CMR-LGE} images,'' {\em Magnetic Resonance Materials in Physics, Biology and Medicine}, vol.~32, pp.~187--195, 2019.

\bibitem{journal/CDHJ/popescu2022}
D.~M. Popescu, H.~G. Abramson, {\em et~al.}, ``Anatomically informed deep learning on contrast-enhanced cardiac magnetic resonance imaging for scar segmentation and clinical feature extraction,'' {\em Cardiovascular Digital Health Journal}, vol.~3, no.~1, pp.~2--13, 2022.

\bibitem{journal/JC/wieslander2013}
B.~Wieslander, K.~C. Wu, {\em et~al.}, ``Localization of myocardial scar in patients with cardiomyopathy and left bundle branch block using electrocardiographic {S}elvester {QRS} scoring,'' {\em Journal of electrocardiology}, vol.~46, no.~3, pp.~249--255, 2013.

\bibitem{journal/PRL/baloglu2019}
U.~B. Baloglu, M.~Talo, {\em et~al.}, ``Classification of myocardial infarction with multi-lead {ECG} signals and deep {CNN},'' {\em Pattern recognition letters}, vol.~122, pp.~23--30, 2019.

\bibitem{journal/CMPB/xiong2021}
P.~Xiong, Y.~Xue, {\em et~al.}, ``Localization of myocardial infarction with multi-lead {ECG} based on densenet,'' {\em Computer Methods and Programs in Biomedicine}, vol.~203, p.~106024, 2021.

\bibitem{journal/TMI/ghimire2019}
S.~Ghimire, J.~L. Sapp, B.~M. Hor{\'a}{\v{c}}ek, and L.~Wang, ``Noninvasive reconstruction of transmural transmembrane potential with simultaneous estimation of prior model error,'' {\em IEEE Transactions on Medical Imaging}, vol.~38, no.~11, pp.~2582--2595, 2019.

\bibitem{journal/MedIA/li2023survey}
L.~Li, W.~Ding, L.~Huang, X.~Zhuang, and V.~Grau, ``Multi-modality cardiac image computing: A survey,'' {\em Medical Image Analysis}, p.~102869, 2023.

\bibitem{journal/PRO/brett2021}
C.~L. Brett, J.~A. Cook, {\em et~al.}, ``Novel workflow for conversion of catheter-based electroanatomic mapping to {DICOM} imaging for noninvasive radioablation of ventricular tachycardia,'' {\em Practical Radiation Oncology}, vol.~11, no.~1, pp.~84--88, 2021.

\bibitem{conf/ICML/radford2021}
A.~Radford, J.~W. Kim, {\em et~al.}, ``Learning transferable visual models from natural language supervision,'' in {\em International conference on machine learning}, pp.~8748--8763, PMLR, 2021.

\bibitem{conf/ICML/jia2021}
C.~Jia, Y.~Yang, {\em et~al.}, ``Scaling up visual and vision-language representation learning with noisy text supervision,'' in {\em International conference on machine learning}, pp.~4904--4916, PMLR, 2021.

\bibitem{conf/STACOM/meister2020}
F.~Meister {\em et~al.}, ``Graph convolutional regression of cardiac depolarization from sparse endocardial maps,'' in {\em Statistical Atlases and Computational Models of the Heart}, pp.~23--34, 2020.

\bibitem{conf/WWWC/khattar2019}
D.~Khattar, J.~S. Goud, M.~Gupta, and V.~Varma, ``{MVAE}: Multimodal variational autoencoder for fake news detection,'' in {\em The world wide web conference}, pp.~2915--2921, 2019.

\bibitem{conf/MICCAI/aguila2023}
A.~L. Aguila, J.~Chapman, and A.~Altmann, ``Multi-modal variational autoencoders for normative modelling across multiple imaging modalities,'' {\em arXiv preprint arXiv:2303.12706}, 2023.

\bibitem{conf/STACOM/li2023}
L.~Li, J.~Camps, {\em et~al.}, ``Deep computational model for the inference of ventricular activation properties,'' in {\em Statistical Atlases and Computational Models of the Heart}, pp.~369--380, 2023.

\bibitem{book/SSBM/kaipio2006}
J.~Kaipio and E.~Somersalo, {\em Statistical and computational inverse problems}, vol.~160.
\newblock Springer Science \& Business Media, 2006.

\bibitem{journal/TMI/yu2015}
L.~Yu, Z.~Zhou, and B.~He, ``Temporal sparse promoting three dimensional imaging of cardiac activation,'' {\em IEEE Transactions on Medical Imaging}, vol.~34, no.~11, pp.~2309--2319, 2015.

\bibitem{journal/FP/karoui2018}
A.~Karoui, L.~Bear, P.~Migerditichan, and N.~Zemzemi, ``Evaluation of fifteen algorithms for the resolution of the electrocardiography imaging inverse problem using ex-vivo and in-silico data,'' {\em Frontiers in Physiology}, vol.~9, p.~1708, 2018.

\bibitem{conf/MICCAI/xu2015}
J.~Xu, J.~L. Sapp, {\em et~al.}, ``Robust transmural electrophysiological imaging: Integrating sparse and dynamic physiological models into {ECG}-based inference,'' in {\em Medical Image Computing and Computer-Assisted Intervention}, pp.~519--527, 2015.

\bibitem{journal/MedIA/camps2021}
J.~Camps, B.~Lawson, {\em et~al.}, ``Inference of ventricular activation properties from non-invasive electrocardiography,'' {\em Medical Image Analysis}, vol.~73, p.~102143, 2021.

\bibitem{journal/EP/bacoyannis2021}
T.~Bacoyannis {\em et~al.}, ``Deep learning formulation of electrocardiographic imaging integrating image and signal information with data-driven regularization,'' {\em EP Europace}, vol.~23, pp.~i55--i62, 2021.

\bibitem{journal/TBME/schuler2021}
S.~Schuler, M.~Schaufelberger, {\em et~al.}, ``Reducing line-of-block artifacts in cardiac activation maps estimated using {ECG} imaging: A comparison of source models and estimation methods,'' {\em IEEE Transactions on Biomedical Engineering}, vol.~69, no.~6, pp.~2041--2052, 2021.

\bibitem{journal/FiP/kalinin2019}
A.~Kalinin, D.~Potyagaylo, and V.~Kalinin, ``Solving the inverse problem of electrocardiography on the endocardium using a single layer source,'' {\em Frontiers in physiology}, p.~58, 2019.

\bibitem{journal/AJPHCP/zhang2005}
X.~Zhang, I.~Ramachandra, {\em et~al.}, ``Noninvasive three-dimensional electrocardiographic imaging of ventricular activation sequence,'' {\em American Journal of Physiology-Heart and Circulatory Physiology}, vol.~289, no.~6, pp.~H2724--H2732, 2005.

\bibitem{journal/MBEC/cluitmans2018}
M.~Cluitmans {\em et~al.}, ``Wavelet-promoted sparsity for non-invasive reconstruction of electrical activity of the heart,'' {\em Medical \& biological engineering \& computing}, vol.~56, pp.~2039--2050, 2018.

\bibitem{journal/TMI/jiang2022}
X.~Jiang, M.~Toloubidokhti, {\em et~al.}, ``Improving generalization by learning geometry-dependent and physics-based reconstruction of image sequences,'' {\em IEEE Transactions on Medical Imaging}, vol.~42, no.~2, pp.~403--415, 2022.

\bibitem{conf/STACOM/meister2021}
F.~Meister, T.~Passerini, {\em et~al.}, ``Graph convolutional regression of cardiac depolarization from sparse endocardial maps,'' in {\em Statistical Atlases and Computational Models of the Heart. M\&Ms and EMIDEC Challenges: 11th International Workshop, STACOM 2020, Held in Conjunction with MICCAI 2020, Lima, Peru, October 4, 2020, Revised Selected Papers 11}, pp.~23--34, Springer, 2021.

\bibitem{journal/Sensors/chen2022}
K.-W. Chen, L.~Bear, and C.-W. Lin, ``Solving inverse electrocardiographic mapping using machine learning and deep learning frameworks,'' {\em Sensors}, vol.~22, no.~6, p.~2331, 2022.

\bibitem{journal/EP/pezzuto2021}
S.~Pezzuto, F.~W. Prinzen, M.~Potse, F.~Maffessanti, F.~Regoli, M.~L. Caputo, G.~Conte, R.~Krause, and A.~Auricchio, ``Reconstruction of three-dimensional biventricular activation based on the 12-lead electrocardiogram via patient-specific modelling,'' {\em EP Europace}, vol.~23, no.~4, pp.~640--647, 2021.

\bibitem{conf/MICCAI/xu2014}
J.~Xu, J.~L. Sapp, A.~Rahimi~Dehaghani, F.~Gao, and L.~Wang, ``Variational bayesian electrophysiological imaging of myocardial infarction,'' in {\em Medical Image Computing and Computer-Assisted Intervention--MICCAI 2014: 17th International Conference, Boston, MA, USA, September 14-18, 2014, Proceedings, Part II 17}, pp.~529--537, 2014.

\bibitem{journal/CCR/nikus2014}
K.~Nikus, Y.~Birnbaum, {\em et~al.}, ``Updated electrocardiographic classification of acute coronary syndromes,'' {\em Current Cardiology Reviews}, vol.~10, no.~3, pp.~229--236, 2014.

\bibitem{journal/PTRSA/banerjee2021}
A.~Banerjee, J.~Camps, {\em et~al.}, ``A completely automated pipeline for {3D} reconstruction of human heart from {2D} cine magnetic resonance slices,'' {\em Philosophical Transactions of the Royal Society A}, vol.~379, no.~2212, p.~20200257, 2021.

\bibitem{conf/EMBC/smith2022}
H.~J. Smith, A.~Banerjee, R.~P. Choudhury, and V.~Grau, ``Automated torso contour extraction from clinical cardiac {MR} slices for {3D} torso reconstruction,'' in {\em Annual International Conference of the IEEE Engineering in Medicine \& Biology Society}, pp.~3809--3813, 2022.

\bibitem{journal/MedIA/schuler2021}
S.~Schuler, N.~Pilia, D.~Potyagaylo, and A.~Loewe, ``Cobiveco: Consistent biventricular coordinates for precise and intuitive description of position in the heart--with matlab implementation,'' {\em Medical Image Analysis}, vol.~74, p.~102247, 2021.

\bibitem{journal/EHJ/thygesen2019}
K.~Thygesen, J.~S. Alpert, {\em et~al.}, ``Fourth universal definition of myocardial infarction (2018),'' {\em European Heart Journal}, vol.~40, no.~3, pp.~237--269, 2019.

\bibitem{journal/CR/gima2002}
K.~Gima and Y.~Rudy, ``Ionic current basis of electrocardiographic waveforms: a model study,'' {\em Circulation Research}, vol.~90, no.~8, pp.~889--896, 2002.

\bibitem{journal/FiP/minchole2019}
A.~Minchol{\'e}, E.~Zacur, {\em et~al.}, ``{MRI}-based computational torso/biventricular multiscale models to investigate the impact of anatomical variability on the {ECG QRS} complex,'' {\em Frontiers in Physiology}, vol.~10, p.~1103, 2019.

\bibitem{journal/Circulation/de1993}
J.~De~Bakker, F.~Van~Capelle, {\em et~al.}, ``Slow conduction in the infarcted human heart.{`Zigzag'}course of activation,'' {\em Circulation}, vol.~88, no.~3, pp.~915--926, 1993.

\bibitem{journal/ANIP/qi2017}
C.~R. Qi, L.~Yi, H.~Su, and L.~J. Guibas, ``Pointnet++: Deep hierarchical feature learning on point sets in a metric space,'' {\em Advances in neural information processing systems}, vol.~30, 2017.

\bibitem{journal/MedIA/beetz2023}
M.~Beetz, A.~Banerjee, J.~Ossenberg-Engels, and V.~Grau, ``Multi-class point cloud completion networks for {3D} cardiac anatomy reconstruction from cine magnetic resonance images,'' {\em Medical Image Analysis}, p.~102975, 2023.

\bibitem{journal/SR/zhu2019}
F.~Zhu, F.~Ye, Y.~Fu, Q.~Liu, and B.~Shen, ``Electrocardiogram generation with a bidirectional {LSTM-CNN} generative adversarial network,'' {\em Scientific reports}, vol.~9, no.~1, p.~6734, 2019.

\bibitem{journal/FiP/beetz2022}
M.~Beetz, A.~Banerjee, and V.~Grau, ``Multi-domain variational autoencoders for combined modeling of {MRI}-based biventricular anatomy and {ECG}-based cardiac electrophysiology,'' {\em Frontiers in physiology}, vol.~13, p.~886723, 2022.

\bibitem{journal/EHJ/littlejohns2019}
T.~J. Littlejohns, C.~Sudlow, N.~E. Allen, and R.~Collins, ``{UK B}iobank: opportunities for cardiovascular research,'' {\em European heart journal}, vol.~40, no.~14, pp.~1158--1166, 2019.

\bibitem{journal/EP/cardone2016}
L.~Cardone-Noott, A.~Bueno-Orovio, {\em et~al.}, ``Human ventricular activation sequence and the simulation of the electrocardiographic {QRS} complex and its variability in healthy and intraventricular block conditions,'' {\em EP Europace}, vol.~18, no.~suppl\_4, pp.~iv4--iv15, 2016.

\bibitem{conf/CiC/camps2022}
J.~Camps, Z.~J. Wang, {\em et~al.}, ``Inference of number and location of purkinje root nodes and ventricular conduction properties from clinical 12-lead {ECG}s for cardiac digital twinning,'' in {\em Computing in Cardiology}, vol.~498, pp.~1--4, 2022.

\bibitem{journal/CR/myerburg1978}
R.~J. Myerburg, H.~Gelband, K.~Nilsson, A.~Castellanos, A.~R. Morales, and A.~L. Bassett, ``The role of canine superficial ventricular muscle fibers in endocardial impulse distribution,'' {\em Circulation Research}, vol.~42, no.~1, pp.~27--35, 1978.

\bibitem{journal/JMCC/taggart2000}
P.~Taggart, P.~M. Sutton, T.~Opthof, R.~Coronel, R.~Trimlett, W.~Pugsley, and P.~Kallis, ``Inhomogeneous transmural conduction during early ischaemia in patients with coronary artery disease,'' {\em Journal of Molecular and Cellular Cardiology}, vol.~32, no.~4, pp.~621--630, 2000.

\bibitem{conf/STACOM/beetz2021}
M.~Beetz, A.~Banerjee, and V.~Grau, ``Generating subpopulation-specific biventricular anatomy models using conditional point cloud variational autoencoders,'' in {\em International Workshop on Statistical Atlases and Computational Models of the Heart}, pp.~75--83, Springer, 2021.

\bibitem{journal/CJ/cupa2018}
J.~Cupa, I.~Strebel, {\em et~al.}, ``Diagnostic and prognostic value of {QRS} duration and {QTc} interval in patients with suspected myocardial infarction,'' {\em Cardiology Journal}, vol.~25, no.~5, pp.~601--610, 2018.

\bibitem{journal/JACC/delewi2013}
R.~Delewi, G.~IJff, T.~P. van~de Hoef, {\em et~al.}, ``Pathological {Q} waves in myocardial infarction in patients treated by primary {PCI},'' {\em JACC: Cardiovascular Imaging}, vol.~6, no.~3, pp.~324--331, 2013.

\bibitem{journal/IJC/kurisu2015}
S.~Kurisu, T.~Iwasaki, {\em et~al.}, ``Poor {R}-wave progression and myocardial infarct size after anterior myocardial infarction in the coronary intervention era,'' {\em IJC Heart \& Vasculature}, vol.~7, pp.~106--109, 2015.

\bibitem{journal/Circulation/das2006}
M.~K. Das, B.~Khan, {\em et~al.}, ``Significance of a fragmented {QRS} complex versus a {Q} wave in patients with coronary artery disease,'' {\em Circulation}, vol.~113, no.~21, pp.~2495--2501, 2006.

\bibitem{journal/IJC/mittal1986}
S.~Mittal and P.~Srivastava, ``Differentiation of poor {R} wave progression of old anteroseptal myocardial infarction from that due to emphysema,'' {\em International Journal of Cardiology}, vol.~13, no.~1, pp.~92--94, 1986.

\bibitem{journal/FiP/luo2020}
G.~Luo, Q.~Li, {\em et~al.}, ``The predictive value of fragmented {QRS} for cardiovascular events in acute myocardial infarction: a systematic review and meta-analysis,'' {\em Frontiers in Physiology}, vol.~11, p.~1027, 2020.

\end{thebibliography}

\end{document}